\documentclass[journal,twoside,web]{ieeecolor}
\usepackage{tmi}
\usepackage{cite}
\usepackage{amsmath,amssymb,amsfonts}
\usepackage{algorithmic}
\usepackage{graphicx}
\usepackage{textcomp}
\usepackage{array}
\usepackage{multirow}
\usepackage{diagbox}
\usepackage[T1]{fontenc}
\usepackage[utf8]{inputenc}
\usepackage{authblk}

\if CLASSOPTIONcompsoc
\usepackage[caption=false, font=normalsize, labelfont=sf, textfont=sf]{subfig}
\else
\usepackage[caption=false, font=footnotesize]{subfig}

\usepackage{lipsum,mwe,cuted}
\usepackage{float}
\usepackage{caption}
\usepackage{booktabs}
\newcommand{\tabincell}[2]{\begin{tabular}{@{}#1@{}}#2\end{tabular}}  
\def\BibTeX{{\rm B\kern-.05em{\sc i\kern-.025em b}\kern-.08em
    T\kern-.1667em\lower.7ex\hbox{E}\kern-.125emX}}
\title{Chronological Age Estimation of Lateral Cephalometric Radiographs with Deep Learning}
\author[1,3,4]{Zhi\-yong Zhang\thanks{zzy20011126@mail.xjtu.edu.cn}}
\author[2]{Ning\-tao Liu\thanks{nt\_liu@stu.xidian.edu.cn}}
\author[2]{Shui\-ping Gou\thanks{shpgou@mail.xidian.edu.cn}}
\author[3]{Chun\-xia Yan\thanks{yanchunxia@mail.xjtu.edu.cn}}
\author[5]{Wen\-fan Jing\thanks{dexterjing@163.com}}
\affil[1]{Key Laboratory of Shaanxi Province for Craniofacial Precision Medicine Research, College of Stomatology, Xi’an Jiaotong University Health Science Center, Xi’an 710004, Shaanxi, People’s Republic of China }
\affil[2]{
Key Laboratory of Intelligent Perception and Image Understanding of Ministry of Education, 
School of Artificial Intelligence, Xidian University, Xi’an 710071, People’s Republic of China}
\affil[3]{
    Forensic Medicine, Xi’an Jiaotong University Health Science Center, Xi’an 710061, Shaanxi, People’s Republic of China
}
\affil[4]{
    Department of Orthodontics, the Affiliated Stomatological Hospital
     of Xi’an Jiaotong University Health Science Center, Xi’an 710004, Shaanxi, People’s Republic of China
}
\affil[5]{
    Department of Radiology, the Affiliated Stomatological Hospital of Xi’an Jiaotong University Health 
    Science Center, Xi’an 710004, Shaanxi, People’s Republic of China
}

\begin{document}
\maketitle

\begin{abstract}
Traditional manual age estimation method is crucial labour based on many kinds of X-Ray image. 
Some current studies have shown that lateral cephalometric(LC) images can be used to estimate age.
However, these methods are based on manually measuring some image features and making age estimates based on experience or scoring.
Therefore, these methods are time-consuming and labor-intensive, and the effect will be affected by subjective opinions.
In this work, we propose a saliency map-enhanced age estimation method, 
which can automatically perform age estimation based on LC images.
Meanwhile it can also show the importance of each region in the image for age estimation, 
which undoubtedly increases the method’s Interpretability. 
Our method was tested on 3014 LC images from 4 to 40 years old. 
The MEA of the experimental result is 1.250, which is less than the result of the state-of-the-art benchmark,
because it performs significantly better in the age group with less data.
In addition, our model is trained in each area with high contribution to age estimation in LC images, 
so the effect of these different areas on the age estimation task were verified.
Consequently, we conclude that the proposed saliency map enhancemented chronological age estimation method of lateral cephalometric radiographs can work well in chronological age estimation task, 
especially when the amount of data is small.
In addition, compared with traditional deep learning, our method is also interpretable.

\end{abstract}

\begin{IEEEkeywords}
Enter key words or phrases in alphabetical 
order, separated by commas. For a list of suggested keywords, send a blank 
e-mail to keywords@ieee.org or visit \underline
{http://www.ieee.org/organizations/pubs/ani\_prod/keywrd98.txt}
\end{IEEEkeywords}

\section{Introduction}
\label{sec:introduction}
Age is defined as “the length of time that a person has lived or a thing has existed”, 
it is one of the important factors in determining a person's identity. 
Age estimation is widely used in forensic human identification and criminal and civil proceedings, 
such as: identification of victims after mass disasters, 
juvenile delinquency, 
migrants without identity papers, 
retirement benefits, etc\cite{haines1972dental}.  

Many parts of the human body can be used for age estimation due to aging changes. 
During the course of a human’s skeletal maturation and degeneration, 
age-related morphological changes take place. 
Age can be estimated from the evaluation of the size, 
shape and degree of epiphyseal ossification of bones. 
Bones from various parts of the body are used to study bone age estimate,
such as hand-wrist\cite{cameriere2012accuracy}, 
knee\cite{vieth2018forensic}, 
foot\cite{hackman2013age}, 
clavicle\cite{langley2016lateral}, 
ilium and pubis\cite{savall2018age}. 
Most of the bones development are completed in their 20s, 
There are few studies on the relationship between age-related bone changes and age in adults, 
except for the pubic bone. 
The significant variation of pubic symphysis morphology related to bone degeneration makes the accuracy of age estimation lower, 
especially over 40 years of age\cite{hartnett2010analysis}. 
The pubis can only be used for corpse to estimate age, which limits its use.

Teeth have the advantage of being preserved for a long time after the disintegration of other tissues, 
even bones, 
and unlike bones, 
they can be clinically inspected directly in the living individuals. 
Every tooth has a unique set of features such as shape, 
pathology, 
wear pattern, 
color and location and the arrangement of teeth in different mouths varies from person to person which forms the basis of identification. 
Tooth and dental characteristics are considered to be one of the most valuable personalized characteristics of the human body,
which provide very persuasive evidence for human body identification. 
The principal basis of dental identification is that no two mouths are alike and each person's teeth are unique\cite{krishan2015dental}. 
These benefits make teeth become the preferred organ for forensic age estimation and recognition\cite{sweet2010forensic}.

Many methods of estimate age from teeth have been established, 
these are divided into four categories, 
clinical/visual, morphological, 
radiologic and biochemical methods based on degradation process observed in tooth structure\cite{savita2017teeth}\cite{stavrianos2008dental} 
Tooth development and eruption sequence have been widely used in the age estimation of children and adolescents with high accuracy (standard error$\pm$1–2 years)\cite{anderson1976age}\cite{demirjian1973new}\cite{moorrees1963age}\cite{seth2018dental}. 
However, 
For adults, 
because the development of teeth has been completed, 
age can only be estimated by aging changes such as tooth wear, 
periodontal disease, 
root transparency, 
cementum annulation, 
root resorption, 
root roughness increase, 
color change and secondary dentin deposition\cite{Gustafson1950Age}\cite{maples1978improved}\cite{solheim1993new}(Joshi et al, 2019;). 
There are very large individual differences in tooth wear, 
gingival atrophy and root resorption due to different chewing habits, 
dental hygiene habits and previous oral diseases\cite{ball2002critique}\cite{huttner2009effects}\cite{li1995age}\cite{solheim1992recession}\cite{vlaskalic1998etiology}\cite{yun2007age}. 
The measurement of root transparency and cementum annulation requires complex experimental process on extracted teeth\cite{gupta2017age}\cite{mohan2018age}\cite{singhal2010comparative}\cite{solheim1989dental}\cite{wittwer2012age}. 
The tooth colour is affected by the change after death, 
and it is difficult to obtain objective measurement\cite{lackovic2000tooth}\cite{martin2003objective}. 
The commonly used biochemical methods to estimate age are aspartic acid racemization, 
DNA methylation and radiocarbon analysis\cite{alkass2010age}\cite{bekaert2015improved}\cite{spalding2005age}, 
but all need to extract tooth tissue for experiment. 
However, 
it is important that these invasive methods are not suitable for estimating the age of living person for ethical, 
religious, 
cultural or scientific reasons.

Dental X-ray radiographs is one of the main clinical diagnostic tools. 
The most commonly used in dental clinics are intraoral periapical radiographs, 
dental panoramic radiographs and lateral cephalograms. For children and adolescents, 
the time of the emergency of the tooth in the oral cavity and the tooth calcification are basically observed in radiographs. 
Demirjian introduced the most widely used method, 
which defined 8 stages of crown and root development on 7 permanent teeth\cite{demirjian1973new}. 
Cameriere innovatively assessed age in children based on the correlation between age and measurement of open apices in teeth and obtained good accuracy\cite{cameriere2006age}. 
In adulthood when all permanent teeth have been completely formed, 
radiographic age estimation becomes difficult. 
The accuracy of measuring the secondary dentin on dental X-ray radiographs to estimate age needs to be improved\cite{azevedo2015dental}\cite{mittal2016age}. 
These manual age estimation methods are boring, time-consuming and subjective.

Gender determination is crucial for identification because the number of possible matches is reduced by half. 
Gender determination mainly base on sexual dimorphism. 
Sexual dimorphism refers to the difference in appearance, 
size and structure between male and female at the same age. 
The pelvis and skull are traditional gender indicators, 
and their success rate in gender identification is approximately 100\%\cite{patil2005determination}. 
The accuracy of traditional methods to infer gender by measuring teeth can not meet the requirement of forensic identification\cite{gandhi2017significance}\cite{rajarathnam2016mandibular}. 
These manual age and gender estimation methods are boring, 
time-consuming and subjective. 
Therefore,
a method for automatically estimating age and gender is needed to improve the accuracy and repeatability.

Deep learning can analyze medical images intelligent, precise and quickly, 
it has a profound impact on various fields of medicine. 
Dou proposed an unsupervised domain adaptation framework with adversarial learning for cross-modality medical image segmentations and achieved good results on cardiac segmentations\cite{dou2018unsupervised}. 
Hannun developed a deep neural network (DNN) to classify 12 rhythm classes and proved that an end-to-end deep learning approach can reach high diagnostic performance comparable that of cardiologists\cite{hannun2019cardiologist}. 
Pratt presented a method for identifying the learning features of a CNN and applied it in the severity diagnosis of DR in fundus images, 
which provide a useful tool to determine the relation between deep learning classification models and clinical diagnosis procedures\cite{pratt2019feature}. 
At present, 
there are few studies on inferring age and gender through dental X-ray images deep learning. 
All of these studies are based on panoramic radiographs\cite{de2019forensic}\cite{alkaabi2019evaluation}\cite{houssein2020dental}\cite{ilic2019gender}\cite{kim2019development}\cite{vila2020deep}.

Lateral cephalogram and orthopantomogram (OPG) radiographs are routinely taken for each orthodontic patient for diagnostic and treatment planning purpose. 
Compared with panoramic radiograph, lateral cephalogram contains the entire craniofacial bones and soft tissue. 
Because of the way the lateral cephalogram were taken, the left and right craniofacial bones and teeth overlapped together, 
so the lateral cephalogram can provide more information than panoramic radiograph. 
However, 
no one has done any research to infer age and gender based on deep learning of lateral cephalogram. 

In this paper, 
LC images are used for the first time in fully automatic age estimation. 
We also propose a novel deep learning approach to overcome the issue of insensitivity to changes in samples after adulthood when the aforementioned methods and images are applied to age estimation. 
Our approach can not only obtain accurate age estimates efficiently and conveniently, 
but also has strong interpretability, 
which can verify the experience and age estimation rules in clinical.
\section{MATERIALS AND METHODS}
\label{sec:MATERIALS AND METHODS}
\subsection{Data Set}
We obtained a dataset contain 20174 LC (female and  male) from the database of the Stomatological Hospital of Xi’an Jiaotong University Health Science Center, China, 
the age of these image ranges from 4 to 40-year-old.
All the subjects were divided into 4 groups by 5-year age. 
The subjects' age and gender distribution are shown in Table. \ref{tab all-CNNs}. 
Images with obvious age errors, wrong imaging locations,
and poor image quality in the dataset are excluded. 
The accuracy of the age label is guaranteed by the ID card information. 
The age of the subject was calculated by subtracting the photo date from the date of birth and dividing by $365.25$ (due to leap years) and rounding to the nearest hundredth. 
The bit depth of the images in the dataset is 16 bpp. 
The size of most images is $1536\times2304$.
\begin{figure} 
    \center
	  \subfloat[age less than 0]{
       \includegraphics[width=0.45\linewidth]{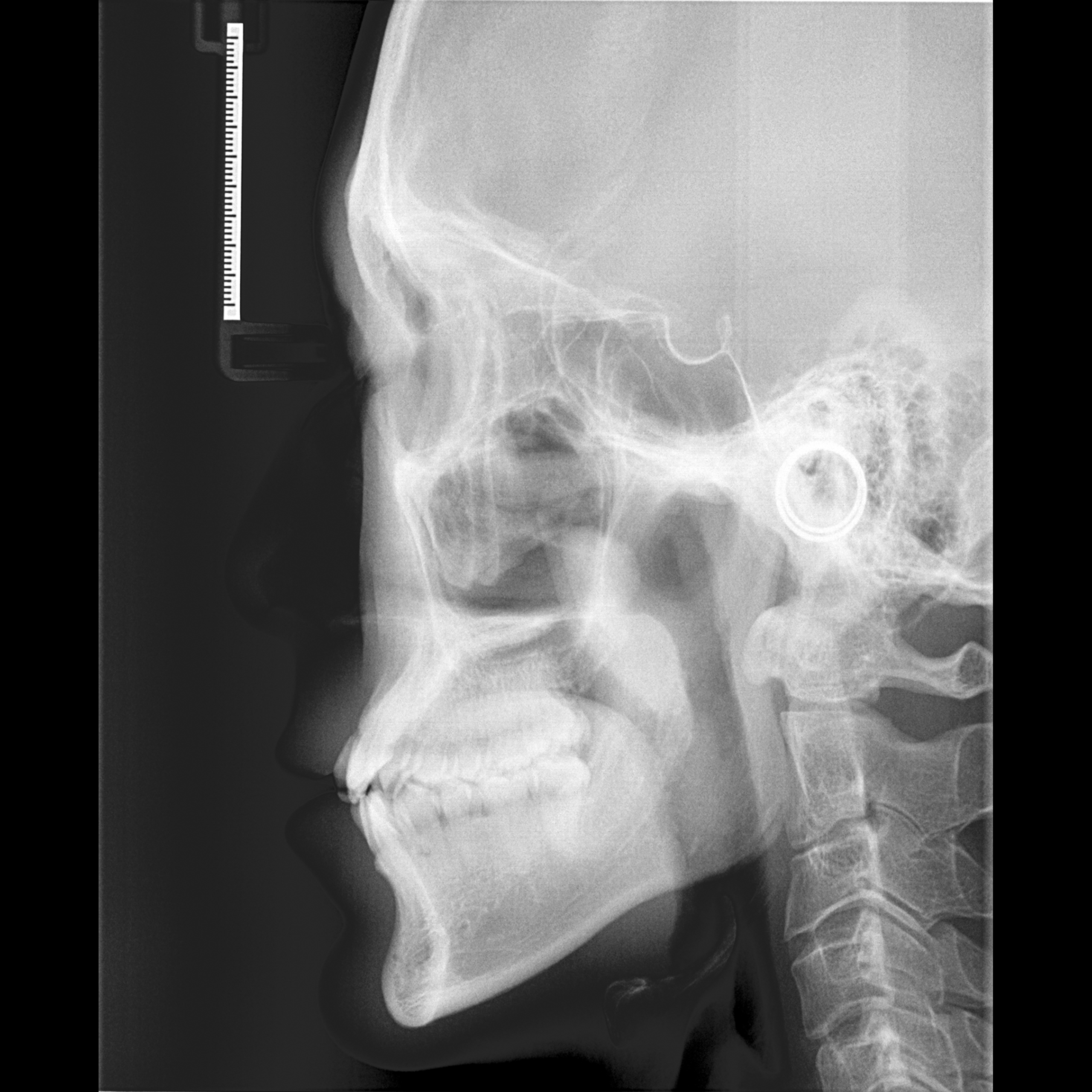}}
    \label{1a}\hfill
	  \subfloat[incomplete]{
        \includegraphics[width=0.45\linewidth]{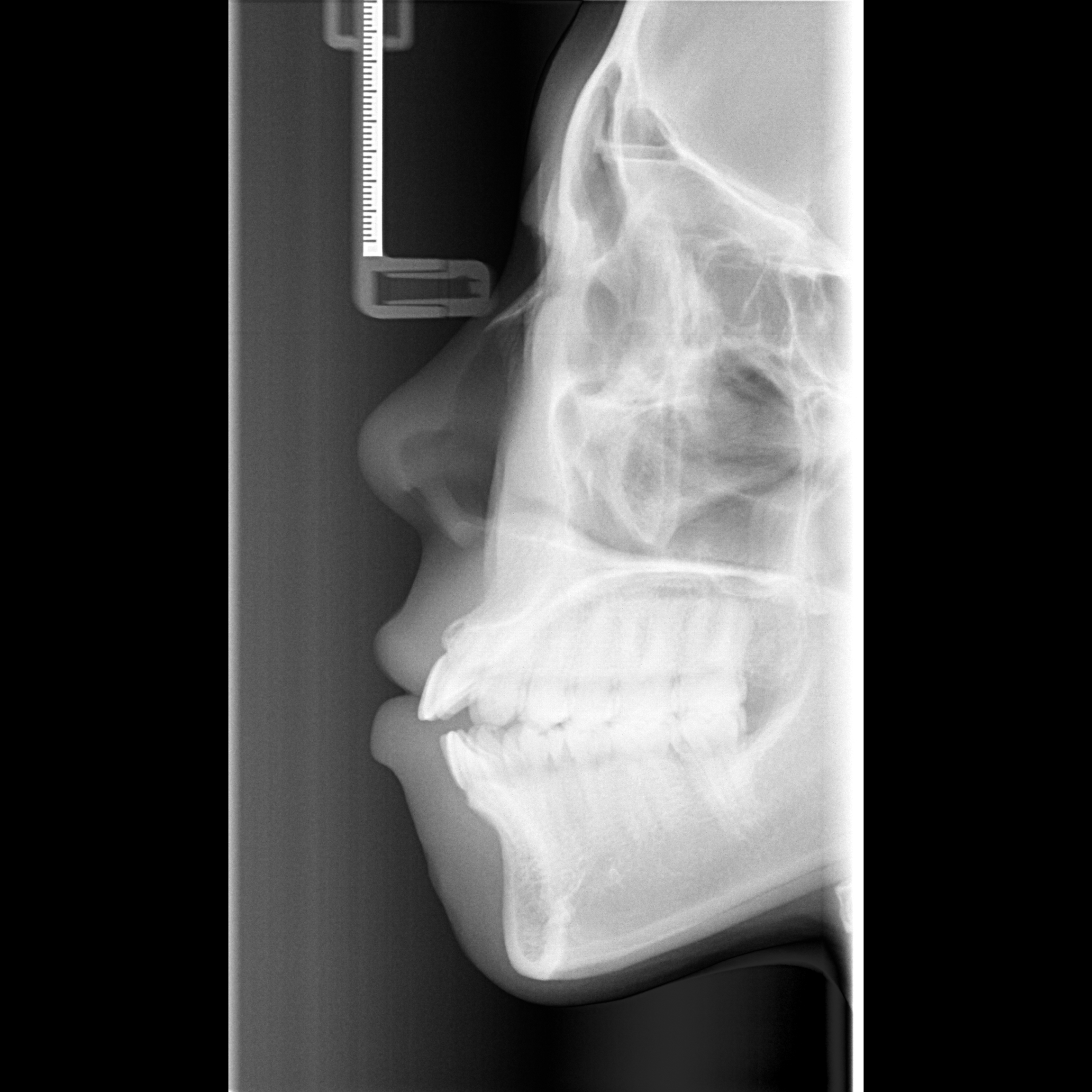}}
    \label{1b}\\
	  \subfloat[with restoration]{
        \includegraphics[width=0.45\linewidth]{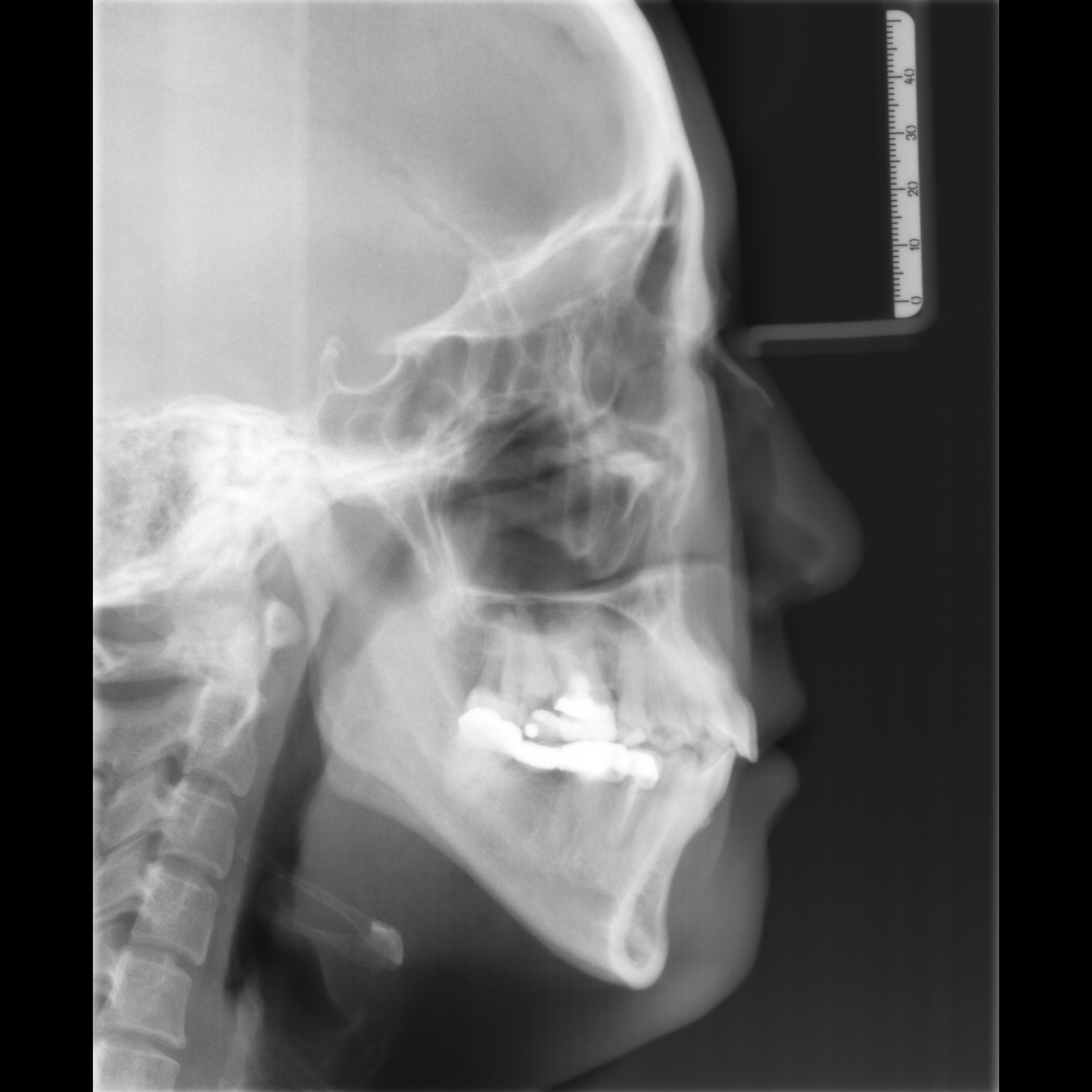}}
    \label{1c}\hfill
	  \subfloat[wrong location]{
        \includegraphics[width=0.45\linewidth]{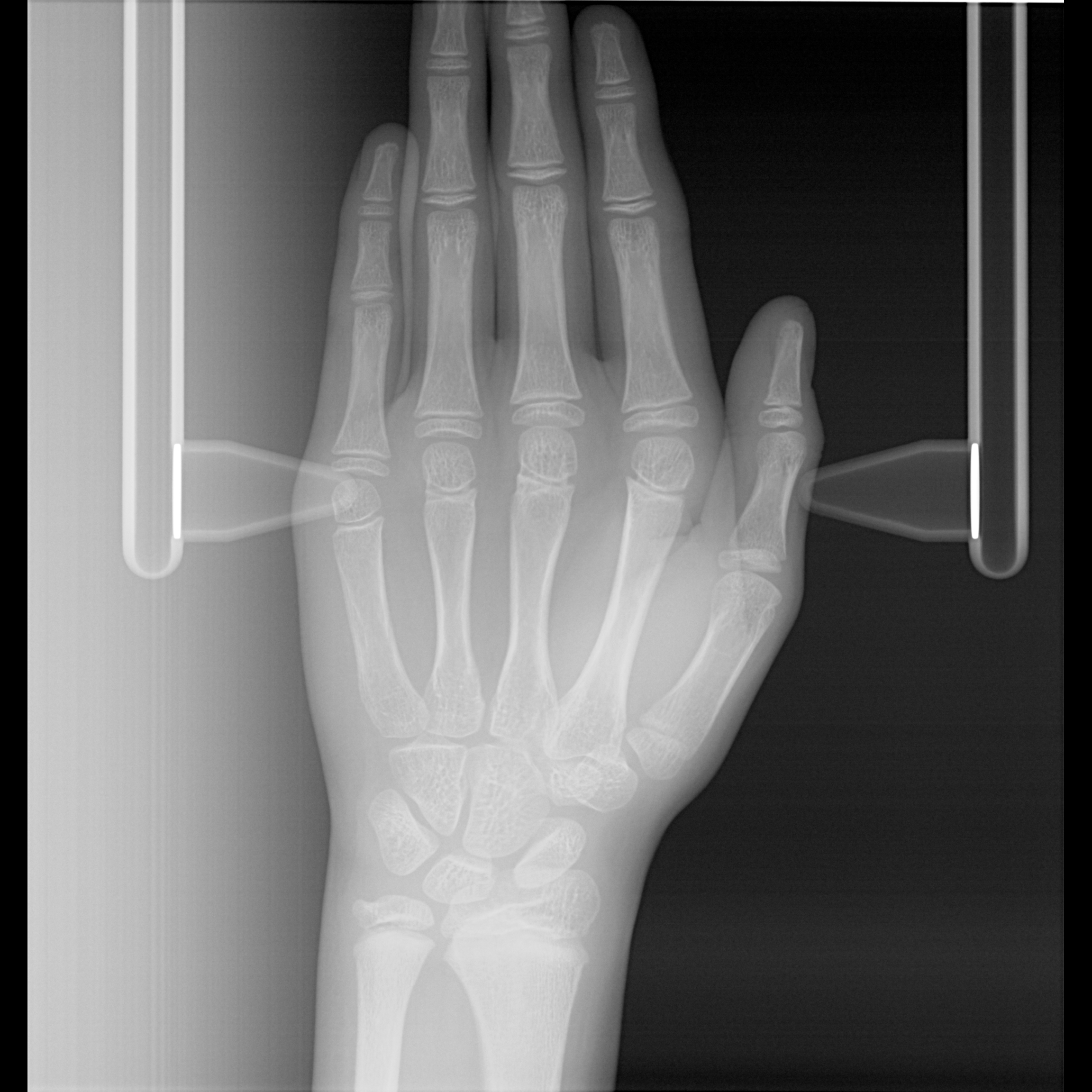}}
     \label{1d} 
	  \caption{Examples of some typical unqualified images}
	  \label{fig unqualified-image} 
\end{figure}

\begin{table}
    \caption{The distribution of the data set}
    \setlength{\tabcolsep}{3pt}
    \begin{tabular}{m{30pt}<{\centering} m{20pt}<{\centering} m{20pt}<{\centering} m{20pt}<{\centering} m{20pt}<{\centering} 
        m{20pt}<{\centering} m{20pt}<{\centering} m{20pt}<{\centering} m{20pt}<{\centering}}
    \toprule[1.5pt]
    \multirow{2}{*}{\tabincell{c}{Age\\(year)}} &
    \multirow{2}{*}{\tabincell{c}{Total\\}} &
    \multicolumn{3}{c}{Data Set}&
    \multicolumn{2}{c}{Gender}&
    \multicolumn{2}{c}{Orientation}\\
    \multicolumn{2}{c}{} & Train & Val & Test & Male & Female & Left & Right\\
    \midrule[0.5pt]
    4-10 & 2599 & 1822 & 393 & 384 & 1264 &	1335 & 2033 & 566\\
    11-15 & 7652 & 5354 & 1148 & 1150 &	3072 & 4580 & 5395 & 2257\\
    16-20 & 4591 & 3211 & 690 &	690 & 1713 & 2878 & 3326 & 1265\\
    21-15 & 3044 & 2137 & 454 & 453 & 815 & 2229 & 2282 & 762\\
    26-30 & 1452 & 1020 & 210 & 222 & 297 & 1155 & 1125 & 327\\
    31-35 & 577 & 408 & 86 & 83 & 102 & 475 & 468 & 109	\\
    36-40 & 259 & 190 & 37 & 32 & 39 & 220 & 203 & 56\\
    All & 20174 & 14142 & 3018 & 3014 & 7302 & 12872 & 14832 & 5342\\
    \bottomrule[1.5pt]
    \end{tabular}
    \label{tab all-CNNs}
\end{table} 

\begin{figure*}[htbp!]
    \centerline{\includegraphics[width=\textwidth]{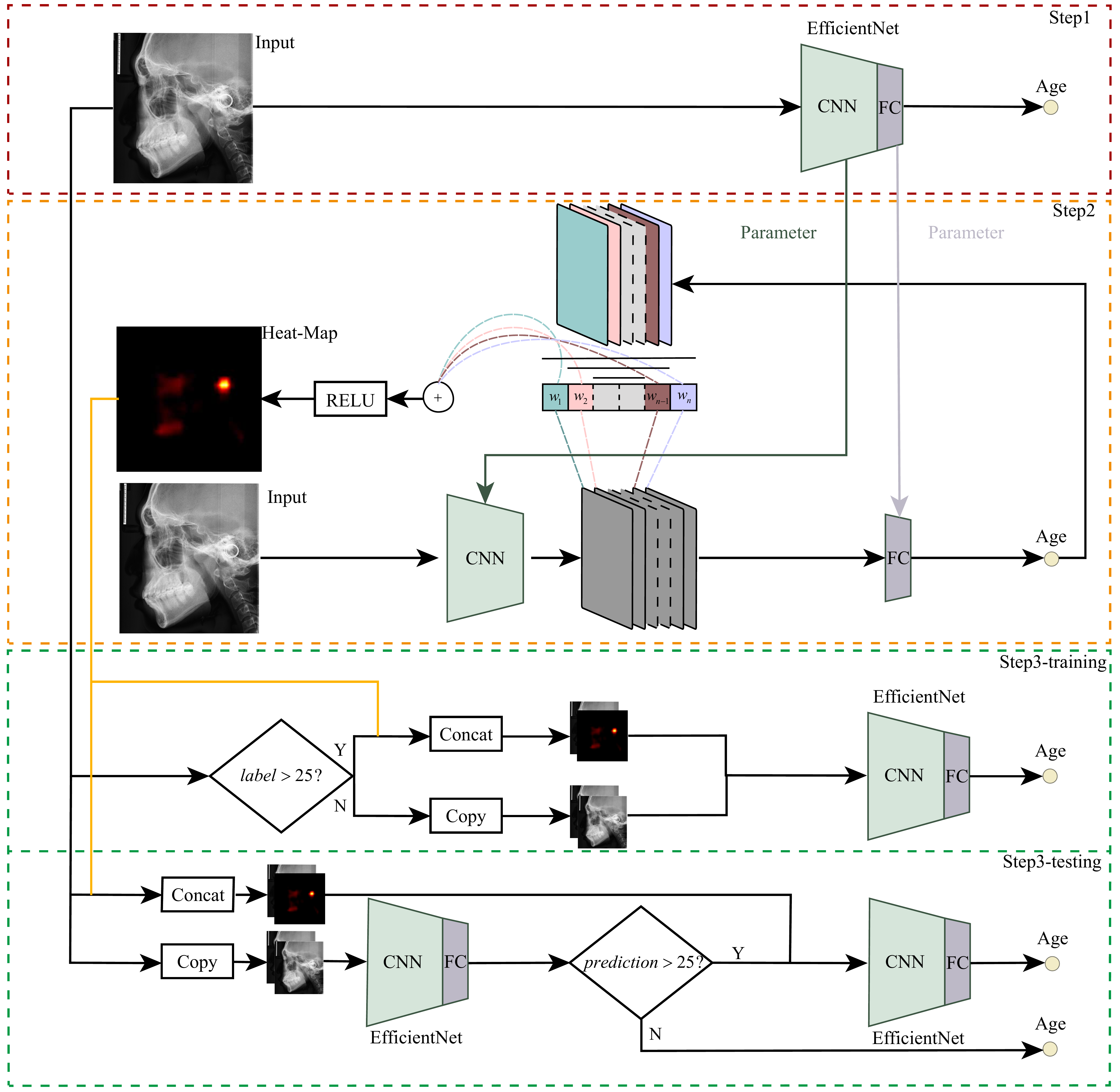}}
    \caption{Step1, Train the Efficient-B0 network, 
    Setp2, Generate each sample's saliency map using the Efficint network trained in Step1.
    Step3-training, 
    in the training phase, 
    because the age label of the LC image is available, 
    if the sample is not more than 25 years old, 
    the LC image and its copy were used as the input of the network; 
    otherwise, 
    the LC image and its corresponding saliency map are used as the input of the network.
    Step3-testing,
    In the testing phase, 
    age labels are not available, 
    and input to the network cannot be determined directly based on age. 
    Therefore, 
    we first used the LC image and its copy as the input of the network. 
    If the estimated age of the network is greater than 25 years old,
    the LC image and its corresponding saliency map were used as the input of the network to estimate the age again, 
    and the latter is used as the final estimated age of the network; 
    otherwise, the output of the network is directly used as the final estimated age without retesting.
    }
    \label{fig model-struct}
\end{figure*}

\subsection{Base Network}
After AlexNet demonstrated its ability in natural image processing in the ImageNet competition in 2012\cite{krizhevsky2017imagenet},
in recent years, 
deep learning algorithms, in particular convolutional networks, 
have rapidly become a methodology of choice for analyzing medical images,
whos application range is very wide,
including different tasks (e.g. segmentation\cite{jha2020doubleu}, classification\cite{esteva2017dermatologist}\cite{setio2016pulmonary}, object detection\cite{han2019synthesizing}, regression\cite{vila2020deep}\cite{lee2017fully}), 
different organs (e.g. head\&neck\cite{zhao2019fully}, lung\cite{shi2020review}, bone\cite{bui2019incorporated}).

Compared with other methods, the most significant features of CNNs are shift invariant and space invariant, 
because the weights in the network are shared by the network performing convolution operations on the image. 
In this way, 
the model does not need to learn separate detectors for the same object occurring in different positions in an image.
A convolutional neural network usually consists of an input and an output layer, 
as well as multiple hidden layers. 
The hidden layers of a CNN typically consist of a series of convolutional layers.
At each convolutional layer $l$, the input image is convolved with a set of kernels with weight
$\mathcal{W}=\lbrace \mathbf{W}_1,  \mathbf{W}_2, \dots , \mathbf{W}_k \rbrace$
and added biases 
$\mathcal{B}=\lbrace b_1, b_2, \dots , b_k \rbrace$,
each generating a new feature map $\mathbf{X}_k$.
The outputs of convolution are subjected to an element-wise no-linear activation function $\sigma\left(\cdot\right)$ like Rectified Linear Unit(ReLU),
$\sigma\left(x\right)=x^{+}=max(0, x)$,
which is subsequently followed by a downsampling operation such as pooling layer.
The same process is repeated for every convolutional layer:
\begin{displaymath}
    \mathbf{X}^l_k=\sigma\left(\mathbf{W}^{l-1}_k\ast\mathbf{X}^{l-1}+b^{l-1}_k\right) \eqno(1)
\end{displaymath}
The parameter sharing of CNN also drastically reduces the amount of parameters
(i.e. the number of weights no longer depends on the size of the input image) that need to be learned\cite{litjens2017survey}.

The hidden layers are considered as a feature extractor, 
it is usually followed by a fully connected layer, 
which takes the feature maps extracted by the hidden layers as input and generate the result.

Backpropagation\cite{hecht1992theory}(BP) is a widely used algorithm for training feedforward neural networks. 
In fitting CNNs, 
backpropagation computes the gradient of the loss function w.r.t the weights and biases of the network for a single input–output example.

Efficient\cite{tan2019efficientnet} is a model that scales the depth, width and resolution dimensions of the CNN collaboratively and uniformly under the condition of fixed computing resources.
In EfficientNet a compound scaling method was proposed, which use a compound coefficient $\phi$ to uniformly scales network width, depth, and resolution in a principled way:
\begin{displaymath}
    \begin{aligned}        
        \text{depth:}\; &d = \alpha^{\phi}\\
        \text{width:}\; &w = \beta^{\phi}\\
        \text{resolution:}\; &r = \gamma^{\phi}\\
        \text{s.t.}\; &\alpha \cdot \beta^{2} \cdot \gamma^{2} \approx 2 \\
        &\alpha \geq 1, \beta \geq 1, \gamma \geq 1 
    \end{aligned}
    \eqno(2)
\end{displaymath}
where $\alpha$, $\beta$, $\beta$ are constants that can be determined by a small grid search. 
Intuitively, $\phi$ is a coefficient describe how many more resources are available for model scaling, 
while $\alpha$, $\beta$, $\gamma$ specify how to assign these extra resources to network width, depth, and resolution respectively.
In this paper, Efficient-B0 is used as our base network.
The main block of Efficient-B0 is mobile inverted bottleneck MBConv\cite{sandler2018mobilenetv2}\cite{tan2019mnasnet}, 
to which the squeeze-and-excitation optimization\cite{hu2018squeeze} is also added. 
The structure of Efficient-B0 is shown in Table \ref{tab2}.
\begin{table}
    \caption{The structure of Efficient-B0}
    \setlength{\tabcolsep}{3pt}
    \begin{tabular}{m{21pt}<{\centering} m{80pt}<{\centering} m{37pt}<{\centering} m{38pt}<{\centering} m{37pt}<{\centering}}
    \toprule[1.5pt]
    \tabincell{c}{Stage\\$i$}& 
    \tabincell{c}{Operator\\$\hat{\mathcal{F}}_{i}$}& 
    \tabincell{c}{Resolution\\$\hat{H}_{i}\times \hat{W}_{i}$}&
    \tabincell{c}{\#Channels\\$\hat{C}_{i}$}&
    \tabincell{c}{\#Layers\\$\hat{L}_{i}$}\\
    \midrule[0.5pt]
    1& Conv$3\times 3$&$224 \times 224$ &32& 1\\
    2& MBConv1,$\text{k}3\times 3$& $112\times 112$ &16& 1\\
    3& MBConv6,$\text{k}3\times 3$& $112\times 112$ &24& 2\\
    4& MBConv6,$\text{k}5\times 5$& $56\times 56$ &24& 2\\
    5& MBConv6,$\text{k}3\times 3$& $28\times 28$ &80& 3\\
    6& MBConv6,$\text{k}5\times 5$& $14\times 14$ &112& 3\\
    7& MBConv6,$\text{k}5\times 5$& $14\times 14$ &192& 4\\
    8& MBConv6,$\text{k}3\times 3$& $7\times 7$ &320& 1\\
    9& Conv$1 \times 1$ \& Pooling \& FC & $7 \times 7$ &1280& 1\\
    \bottomrule[1.5pt]
    \end{tabular}
    \label{tab2}
\end{table}
\subsection{Saliency Map}
Compared with traditional methods, 
CNNs have significant advantages in addressing medical and health care problems. 
But this also brings some serious challenges, 
the most important of which is model interpretability and explainability\cite{vellido2019importance}, especially in medical and health care.
The visualization of the intermediate process and results of the model is common to improve the interpretability of the model.

In this paper, we used Grad-CAM\cite{selvaraju2017grad} as visualization approach without modifying the structure of base network.
in which, the global average of the gradient of the feature map is calculated as its weight.
The weight of the $k-th$ feature map input to the fully connected layer in the base network to the output
\begin{displaymath}
    \alpha_k = \frac{1}{Z}\sum_i\sum_j\frac{\partial \hat{y}}{\partial A^{k}_{ij}} \eqno(3)
\end{displaymath}
where, $Z$ is the feature map, $\hat{y}$ is the output of network, and $A^{k}_{ij}$ is the value of pixel $(i, j)$ in the $k-th$ feature map.
saliency map $M$ is obtained by calculating the weighted sum of the feature map:
\begin{displaymath}
    M = ReLU(\sum_k\alpha_k A^k) \eqno(4)
\end{displaymath}
The process of generating saliency map is shown in the Step2 of Fig.\ref{fig model-struct}

\subsection{Age Estimation with Saliency Map}
Sample imbalance is common in many tasks. In the LC images age estamation task,
the distribution of the count of samples of each chronological age varies greatly, which is shown in Table. \ref{tab all-CNNs}.
This imbalance is also reflected in the performance of using the base model to estimate the age of LC images of each chronological age.

In order to alleviate the problem of data imbalance and improve the overall performance of the model, 
we have considered how to apply the knowledge learned by the model on all training data, 
especially in the age ranges with samall count of samples,
to the samples in those age groups with large number of samples.
Inspired by the positioning ability shown when CAM is applied to Weakly-supervised Object Localization and Fine-grained Recognition in \cite{Zhou_2016_CVPR}, 
we believe that the saliency map generated by Grad-CAM based on the trained model contains knowledge from all training samples, 
which is attention for samples of age ranges with samall count of samples.

In addition, 
in practical applications, 
the interpretability of the method is crucial. 
Users need to know which parts of the LC image are derived from the age estimation results and the contribution of each part.
saliency map can well describe the contribution of each area in the image to the output result and visualize it, 
which is of great significance to our understanding of the working mechanism of the network.
Therefore, 
in this paper, 
the saliency map of each age sample was analyzed, 
and whether it is consistent with the proven age-related changes was also verified.

We compared the performance of models that use LC images as input and LC images and corresponding saliency map generated by Grad-CAM as input. 
We observed that although the overall performance of the model trained with saliency map on the test set is not as good as the model trained with LC Images alone, 
the former performs significantly better than the latter in those age groups with sparse samples, which is shown in Fig.\ref{fig compare-CAM}.

As we have observed, between the ages of 4 and 25, 
when only LC images are used as the input to the network, 
the MAE is less than that of using both LC images and corresponding saliency map as input, 
while between the ages of 26 and 30, 
which is also the age range that traditional method can hardly estimate the age (add a cite),  
using both of LC images and corresponding saliency map as input of network performs better.

Therefore, 
in the train, 
we simply copy the LC image as input for samples from 4 to 25 years old, 
while for the samples from 26 to 40 years old, 
we concated the LC image and corresponding saliency map as input,
because of the age label is available.

However, 
in the test, 
because the age label is not available, 
we applied a different strategy, 
that is,
first just copy the LC image as input to get the predicted age. 
If the predicted age is greater than 25, 
then concat the LC image and correspondingsaliency map as the input of the network, 
test again, 
and the result of the retest is regarded as the final estimated age, 
otherwise, 
the result of the first test is directly used as the final estimated age. 
We call this strategy \textit{retest}
\subsection{Regional comparison}
\begin{figure}[!t]
    \centerline{\includegraphics[width=\columnwidth]{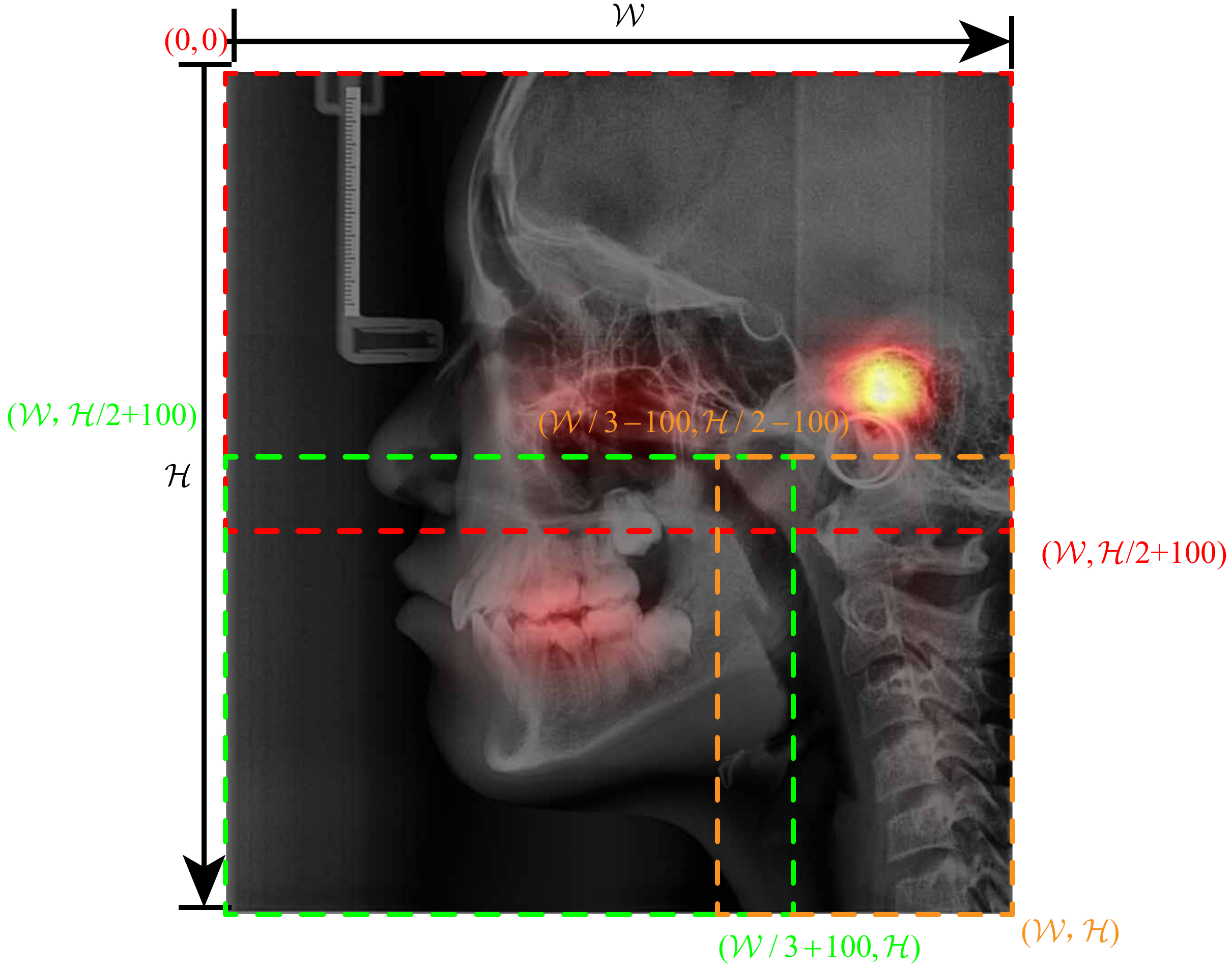}}
    \caption{The green, red and orange box are the boundaries of A, B and C respectively. 
    The coordinates of the upper left and lower right corners of each box are also displayed in corresponding colors}
    \label{fig part-box}
\end{figure}

In the practice of forensic, 
it is very common that all parts of the LC image cannot be obtained.
Therefore, it is necessary to measure the effect of each area in the LC image on the age estimation.
First, calculate the mean saliency map for each age  as follow:
\begin{displaymath}
    M_i = \frac{1}{N_i}\sum^{N_i}_{n=1}m_n \eqno(5)
\end{displaymath}
where $M_i$ is the mean saliency map of age $i$, 
$i\in \lbrace 4, 5, \dots, 39, 40\rbrace$, $N_i$ is the count of samples with age $i$ in training set,
, $m_n$ is the saliency map of $n_{th}$ sample and $\sum$ means element-wise summation.
As the mean saliency map shown in \ref{fig cam-mean-whole},
we can find that the more significant parts in the LC image are mainly concentrated in three parts.
Each LC image was divided into these three overlapping parts with the same strategy, 
because when the LC image is imaged, 
the position and posture of the subject relative to the shooting device are almost fixed.
As shown in Fig. \ref{fig part-box},
for a specific LC image, 
set its width and height as $\mathcal{W}$ and $\mathcal{H}$ respectively, 
let the upper left corner of the LC image as the origin of the coordinates, 
right and down as the positive direction of the x-axis and y-axis, 
and the width of one pixel as the unit length,
establish a coordinate system, 
the coordinates of the upper left and lower right corners of the rectangle of the skull are $(0, 0)$ and $(\mathcal{W}, \frac{1}{2}\mathcal{H}+100)$ respectively.
If the orientation is left, 
the coordinates of the upper left and lower right corners of the rectangle of the tooth part are $(0, \frac{1}{2}\mathcal{H}-100)$, $(\frac{2}{3}\mathcal{W}+100, \mathcal{H})$, 
the upper left and lower right corners of the rectangle of the spine part The coordinates are $(\frac{2}{3}\mathcal{W}-100, \frac{1}{2}\mathcal{H}-100)$, $(\mathcal{(W}, \mathcal{H})$; 
If the orientation is right, 
each part is mirror-symmetrical to those of the image facing the left.

In order to compare their performance in the age estimation and provide guidance for forensic practice, 
each part of the sample was used as the input of the model. 
In order to analyze the contribution of each region in the local image to the age estimation more accurately and verify the reliability of the saliency area of the saliency map of the entire image, 
the saliency maps of each local image were also generated.
\begin{table}[htbp]
    \centering
    \caption{Take the upper left corner of the LC image facing left as the origin of the coordinates, 
    right and down are the positive directions of the x-axis and the y-axis, 
    and the width of a single pixel as the unit length to establish a coordinate system,
    the coordinates $(x,y)$ of the upper left and lower right corners of each part. 
    The boxes of the LC image facing the right are mirror-symmetrical with those of LC image facing the left.
    POS.:Position; ORI.:Orientation}
      \begin{tabular}{ccc}
        \toprule[1.5pt]
        \diagbox[width=40pt,trim=l]{PAR.}{POS.} & Upper Left & Lower Right\\
        \midrule[1pt]
        A & $(0, \frac{1}{2}\mathcal{H}-100)$ & $(\frac{2}{3}\mathcal{W}+100, \mathcal{H})$\\[6pt]
        B & $(0, 0)$ & $(\mathcal{W}, \frac{1}{2}\mathcal{H}+100)$ \\[6pt]
        C & $(\frac{2}{3}\mathcal{W}-100, \frac{1}{2}\mathcal{H}-100)$ & $(\mathcal{(W}, \mathcal{H})$\\[6pt]
        \bottomrule[1.5pt]
      \end{tabular}%
    \label{tab:addlabel}%
\end{table}%

\subsection{Age Estimation with Gender Classification}
\begin{figure}[!t]
    \centerline{\includegraphics[width=\columnwidth]{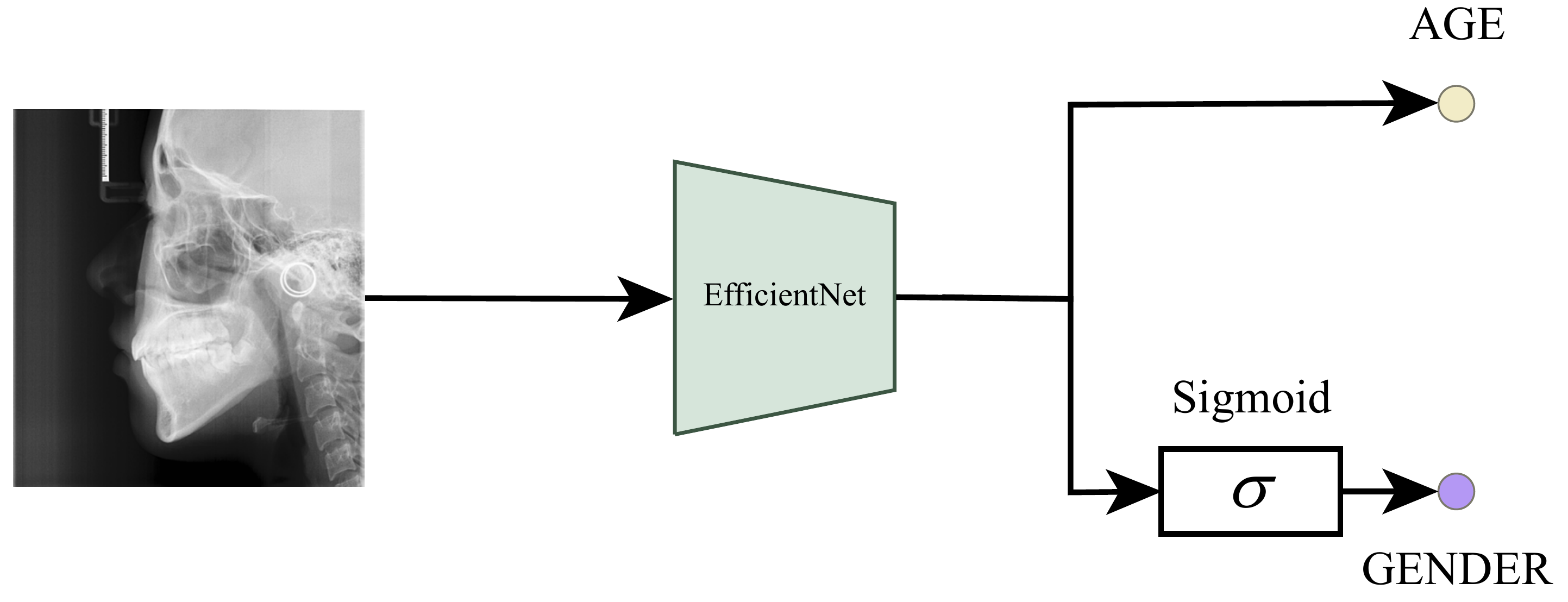}}
    \caption{The output of EfficientNet=B0 includes age and gender, 
    and the gender output needs to be mapped to a value between 0 and 1 through the Sigmoid function. 
    If this value is closer to 0, 
    the gender of the input image is considered to be male, 
    and if it is closer to 1, 
    it is considered to be female (because the label value of male is 0 and female is 1)}
    \label{fig model-estimate-classify}
\end{figure}

Due to differences in male and female development, 
LC images of the same age are bound to be different, which is especially obvious before adulthood.
The age labels are the same, but the characteristics of the images are different, 
or the characteristics are similar, but the age labels are different. 
This inconsistency caused by gender will inevitably make the neural network optimize in different directions.
Therefore, 
adding a gender estimation task to age estimation enables the neural network to distinguish between genders, 
thereby achieving better age estimation performance.
At the same time, gender classification is also crucial for forensic practice.
so we modified the output and loss function of the Efficient-B0 model so that the model has the ability to estimate age and classify gender at the same time,
which is shown in Fig. \ref{fig model-estimate-classify}.

\section{EXPERIMENTS AND PERFORMANCE EVALUATION}
The samples of each age are split into training set, validation set and test set according to the ratio of $7:1.5:1.5$,
which makes each set contain samples of each age proportionally, 
thus avoiding the distribution ratio of samples in each age is inconsistent with the overall distribution ratio.

The weight of the model are randomly initialized with a normal distribution.
In this paper, the model are trained using mini-batch, that is,
in each training iteration, $batch-size$ samples are input to the model, 
and the error of the predicted value is calculated by the loss function.
In our approach, 
the loss functions for age estimation and gender classification are $\mathcal{L}_{age}$ and $\mathcal{L}_{sex}$ respectively, 
which are defined as follows:
\begin{displaymath}
    \mathcal{L}_{age} = \frac{1}{n}\sum^{n}_{i=1}|\hat{y}_a - y_a| \eqno(6)
\end{displaymath}

\begin{displaymath}
    \mathcal{L}_{gender} = \sqrt{\frac{1}{n}\sum^{n}_{i=1}\left(\hat{y}_g - y_g\right)^2} \eqno(7)
\end{displaymath}
where $n$ is the number of samples in a batch (also called batch size), 
$y_a$ and $y_g$ are the age label and gender label of LC image, respectively, 
$\hat{y}_a$ and $\hat{y}_g$ are the age prediction and gender prediction of network, respectively,.

The loss function will be propagated back through the network by BP mentioned in \ref{sec:MATERIALS AND METHODS}, 
and the gradient of each parameter in the network w.r.t the loss of the current training iteration is calculated.
The optimizer used in our work is Adam\cite{kingma2014adam}, and its weight decay is set to $0.0001$.
Different from the traditional stochastic gradient descent, 
Adam designs independent adaptive learning rates for different parameters by calculating the first and second moment estimates of the gradient.

In order to prevent the model from converging to the local optimum and speed up the convergence speed,
the learning rate will also decrease with the number of times the network traverse the whole dataset (also called epoch). 
Suppose the initial learning rate is $\eta$, 
we get the learning rate of the current epoch. 
\begin{displaymath}
    \eta_i = \eta \times 0.8^{\lfloor epoch/4 \rfloor}  \eqno(8)
\end{displaymath}
where the initial learning rate $\eta=0.0001$ and $\lfloor \cdot \rfloor$ means floor operation. 
Each parameter will be updated according to its gradient and current training iteration learning rate:
\begin{displaymath}
    w=w-\eta_i\bigtriangledown_w \eqno(6)
\end{displaymath}
where $w$ is the parameter of model and $\bigtriangledown_w$ is the gradient of parameter $w$ w.r.t the loss.
The process of parameter update is also called \textit{gradient descent}.

In this paper, several typical CNNs that perform well on natural images were tested using the same data set.
For every network, 
when the performance on validation set does not improve for three consecutive epochs or the epoch reaches the max epoch, which was set at $100$ in this paper, 
the training stops. 
After training, 
select the parameters of the network that performs best on the validation set and test on the test set to get the final performance of the model.

For a specific Convolutional Neural Network (CNN) with a Fully Connected layer(FC layer) as a classifier,   
since the feature map is flattened into a one-dimensional vector before being input to the fully connected layer, 
and the input size of the fully connected layer is fixed,
this requires The size of the image input to a convolutional neural network must be fixed(The Global Average Pooling(GAP) proposed in \cite{lin2013network} can free CNNs from this limitation).
Meanwhile, 
considering the preservation of the texture and structure information in the image and the efficiency of network,
the images in the data set are resized to $1000 \times 1000 $.
However, the aspect ratio of the samples in the data set is not $1:1$, 
directly resizing to $1000\times1000$ will inevitably cause image distortion.
Therefore, before resizing
the short side of every LC image in dataset was padded to the length of its long side, 
so that the aspect ratio of the image is $1:1$.
Finally, the LC images are normalized to $\left[0,1\right]$.

The samples in the training set have been augmented to increase the diversity of the samples, 
thereby improving the generalization performance of the model.
The training set are augmented by random affine transformation, 
horizontal flip, and vertical flip.

In order to compare the performance of these models comprehensively, 
several metrics that are meaningful to practic are calculated.
The error of a individual sample(E) is the true age of the sample minus the predicted age of the sample by the network.
The absolute error (AE) of a individual sample is the absolute value of its error.
The median of the errors of all samples is E.Med, and the median of the absolute errors is AE.Med.
The average($\mu$), standard deviation($\sigma$) and interquartile range (IQR) of AE are also calculated,
The mean and median are used to measure the overall age-estamation performance of the network, 
while the stability and availability of the age-estamation performance are measured by the standard deviation and IQR.

There are also several metrics calculated to evaluate the performance of the model on gender-classification.
The output of the neural network on gender classification is a continuous value of 0-1. 
In this paper, 
the threshold was set to $0.5$.
If the output value of network was less than the threshold, 
the predicted result was regarded as male, 
otherwise,
the predicted result was regarded as female
(if the gender of the sample is male, the gender label is coded as 0, otherwise it is coded as 1).
Accuracy is the ratio of the predicted result is consistent with the label, 
which is used to assess whether the network can classify the two categories well.
However, the calculation of accuracy needs to rely on a specific thresholds,
which is not objective and incomplete for network performance evaluation.
Therefore, another metric, ROC-AUC, 
which is widely used to assess the robustness of classification models,
was also calculated. 
When a threshold is set, 
a set of sensitivity and specificitiy will be calculated, 
and a point will be obtained on the plane with sensitivity and specificitiy as the horizontal and vertical coordinates.
ROC is a curve obtained by connecting all of these points, 
and the area under the curve is ROC-AUC.

In order to select a basic network suitable for age estimation and gender classification of LC images
several typical CNNs that perform well on natural images are tested using our data set,
because natural images and medical images have some of the same low-level features, such as texture and edges, etc. 
As shown in Table. \ref{tab all-CNNs}, the EfficientNet-B0 performed much better than other CNNs, 
so it was chosen as our basic network in this work.

After the basic model was trained, 
saliency map is generated to observe the significance of each part in the LC image for age-estimation,
which are shown in Fig.\ref{fig model-struct}. 

By comparing the age estimation performance on test set of the basic network with or without Grad-CAM, 
we built a decision block to determine at which ages need to connect to Grad-CAM and which ages only need to copy LC images,
which are shown in Fig. \ref{fig model-struct}.
Based on this decision block, 
we trained Efficient-B0 again. 
The difference is that during training, 
the decision block can determine whether the input of the network is the concatenation of LC image and Grad-CAM or the LC image and its copy, 
while during testing, 
for all ages, 
the LC image and its Grad-CAM are concatenated as the input of the network.

In order to assess the contribution of each salient part in the LC image, 
we divided the LC image into three parts as shown in Fig. \ref{fig part-box}.
A, B, and C part of  LC image were used as the input of the network respectively to assess the performance of every single part for age estimation.
After that, the A, B, and C parts of the LC image were combined in pairs as the input of network to assess the performance of age estimation when each area of the LC image is missing.
The split of data set, training parameters and network remain unchanged.

After the contributions of Grad-CAM and each part of LC image to age estimation were assessed, 
we assessed the influence of gender classification on age estimation.
When the network performs age estimation and gender classification at the same time, 
instead of simply adding $\mathcal{L}_{age}$ and $\mathcal{L}_{gender}$ to get the overall loss $\mathcal{L}$ (Eq. 7),
$\mathcal{L}_{age}$ and $\mathcal{L}_{gender}$ are multiplied by the factor $\alpha$ and $1-\alpha$ respectively before the addition, 
because the scales of $\mathcal{L}_{age}$ and $\mathcal{L}_{gender}$ are different, 
which can avoid the network tending to optimize $\mathcal{L}_{age}$.
The factor $\alpha$ was set to $0.026$ in this paper, 
which is the ratio of the range of gender label values to the range of age label values.
\begin{displaymath}
    \mathcal{L} = \alpha\mathcal{L}_{age} + \left(1-\alpha\right)\mathcal{L}_{gender} \eqno(9)
\end{displaymath}

\begin{table*}[htbp]
    \centering
    \caption{Comparison of CNN model performance. 
    \#Para: Number of model parameters; 
    \#FLOPs: Floating point operation;
    E.Med.: Median of error;
    $\mu\pm\sigma$: Mean $\pm$ Standard deviation;
    AE.Med.: Median absolute error;
    IQR:Interquartile range.
    The error metrics are given in years.
    }
      \begin{tabular}{m{25pt}<{\centering} m{22pt}<{\centering} ccccccccc}
        \toprule[1.5pt]
            &       & Age   & 4-10  & 11-15 & 16-20 & 21-25 & 26-30 & 31-35 & 36-40 & ALL \\
        \midrule[0.5pt]
      \multirow{4}[0]{*}{Res18} & \multirow{2}[0]{*}{\#Para} & E.Med. & -0.13 & -0.48 & -0.92 & -1.07 & 0.18  & 1.80  & 5.26  & -0.23 \\
            &       & $\mu\pm\sigma$ & 0.93$\pm$0.76 & 1.33$\pm$1.27 & 2.03$\pm$1.70 & 2.49$\pm$1.90 & 2.65$\pm$2.05 & 3.76$\pm$3.84 & 5.99$\pm$4.77 & 2.15$\pm$2.37 \\
            & \multirow{2}[0]{*}{\#Flops} & AE.Med. & 0.77  & 0.96  & 1.52  & 2.05  & 2.27  & 2.56  & 5.26  & 1.43 \\
            &       & IQR   & 1.15  & 1.23  & 2.31  & 2.90  & 2.38  & 4.56  & 6.32  & 2.23 \\
        \midrule[0.5pt]
      \multirow{4}[0]{*}{Res34} & \multirow{2}[0]{*}{\#Para} & E.Med. & -0.20 & -0.08 & -0.38 & -0.35 & -0.18 & 1.94  & 3.585 & -0.05 \\
            &       & $\mu\pm\sigma$    & 0.92$\pm$0.85 & 1.18$\pm$0.96 & 1.64$\pm$1.22 & 2.34$\pm$1.70 & 2.28$\pm$1.78 & 2.89$\pm$2.50 & 5.65$\pm$5.68 & 1.86$\pm$2.01 \\
            & \multirow{2}[0]{*}{\#Flops} & AE.Med. & 0.70  & 0.99  & 1.30  & 2.06  & 1.94  & 2.52  & 3.585 & 1.31 \\
            &       & IQR   & 1.02  & 1.12  & 1.65  & 2.27  & 2.16  & 2.32  & 5.61  & 1.96 \\
        \midrule[0.5pt]
      \multirow{4}[0]{*}{Res50} & \multirow{2}[0]{*}{\#Para} & E.Med. & -0.314 & -0.29 & -0.92 & -1.43 & 0.21  & 2.86  & 4.18  & -0.24 \\
            &       & $\mu\pm\sigma$ & 0.99$\pm$0.84 & 1.36$\pm$1.14 & 2.15$\pm$1.86 & 2.72$\pm$1.97 & 2.76$\pm$2.23 & 4.04$\pm$3.70 & 5.37$\pm$4.28 & 2.24$\pm$2.34 \\
            & \multirow{2}[0]{*}{\#Flops} & AE.Med. & 0.80  & 1.02  & 1.64  & 2.40  & 2.37  & 3.24  & 4.18  & 1.55 \\
            &       & IQR   & 1.00  & 1.25  & 2.25  & 2.14  & 2.62  & 4.48  & 3.82  & 2.37 \\
        \midrule[0.5pt]
      \multirow{4}[0]{*}{Res101} & \multirow{2}[0]{*}{\#Para} & E.Med. & -0.20 & -0.13 & -0.85 & -0.64 & 1.16  & 2.71  & 5.64  & -0.02 \\
            &       & $\mu\pm\sigma$ & 1.00$\pm$0.89 & 1.35$\pm$1.33 & 1.96$\pm$1.56 & 2.27$\pm$1.89 & 2.70$\pm$2.03 & 3.87$\pm$3.34 & 6.05$\pm$4.37 & 2.13$\pm$2.32 \\
            & \multirow{2}[0]{*}{\#Flops} & AE.Med. & 0.81  & 1.06  & 1.61  & 1.85  & 2.20  & 2.78  & 5.64  & 1.41 \\
            &       & IQR   & 1.01  & 1.23  & 2.31  & 2.62  & 2.31  & 4.42  & 5.03  & 2.24 \\
        \midrule[0.5pt]
      \multirow{4}[0]{*}{Res152} & \multirow{2}[0]{*}{\#Para} & E.Med. & -0.20 & -0.13 & -0.66 & -1.05 & 0.31  & 2.29  & 5.21  & -0.17 \\
            &       &$\mu\pm\sigma$ & 1.03$\pm$1.00 & 1.38$\pm$1.17 & 1.92$\pm$1.72 & 2.35$\pm$1.80 & 2.45$\pm$1.56 & 3.77$\pm$3.84 & 5.76$\pm$4.32 & 2.09$\pm$2.31 \\
            & \multirow{2}[0]{*}{\#Flops} & AE.Med. & 0.75  & 1.11  & 1.47  & 2.04  & 2.02  & 2.77  & 5.21  & 1.43 \\
            &       & IQR   & 1.14  & 1.35  & 2.08  & 2.56  & 2.39  & 3.59  & 5.25  & 2.22 \\
        \midrule[0.5pt]
      \multirow{4}[0]{*}{Dense121} & \multirow{2}[0]{*}{\#Para} & E.Med. & 0.33  & 0.26  & -0.30 & -0.53 & 0.62  & 2.54  & 5.47  & 0.31 \\
            &       & $\mu\pm\sigma$ & 1.02$\pm$0.78 & 1.39$\pm$1.24 & 2.00$\pm$1.42 & 2.06$\pm$1.58 & 3.05$\pm$2.29 & 3.82$\pm$3.63 & 5.72$\pm$4.29 & 2.15$\pm$2.25 \\
            & \multirow{2}[0]{*}{\#Flops} & AE.Med. & 0.81  & 1.11  & 1.76  & 1.69  & 2.57  & 3.03  & 5.47  & 1.55 \\
            &       & IQR   & 1.06  & 1.47  & 2.11  & 2.15  & 2.74  & 3.26  & 4.51  & 2.21 \\
        \midrule[0.5pt]
      \multirow{4}[0]{*}{Dense161} & \multirow{2}[0]{*}{\#Para} & E.Med. & -0.61 & -0.30 & -0.33 & -0.58 & 1.57  & 4.28  & 7.39  & 0.06 \\
            &       &$\mu\pm\sigma$ & 1.02$\pm$0.80 & 1.15$\pm$0.97 & 1.85$\pm$1.62 & 2.37$\pm$1.92 & 2.71$\pm$1.96 & 4.91$\pm$3.68 & 7.94$\pm$4.53 & 2.27$\pm$2.52 \\
            & \multirow{2}[0]{*}{\#Flops} & AE.Med. & 0.85  & 0.90  & 1.44  & 1.90  & 2.44  & 4.28  & 7.39  & 1.50 \\
            &       & IQR   & 1.13  & 1.18  & 1.73  & 2.65  & 2.53  & 4.15  & 6.24  & 2.36 \\
        \midrule[0.5pt]
      \multirow{4}[0]{*}{IncepV4} & \multirow{2}[0]{*}{\#Para} & E.Med. & -0.27 & -0.30 & -0.75 & -0.37 & 0.72 & 2.68 & 6.71 &  -0.05 \\
            &       &$\mu\pm\sigma$ &0.80$\pm$0.68& 1.16$\pm$12 &1.82$\pm$1.50&2.42$\pm$1.96&2.49$\pm$2.16&3.53$\pm$3.56& 6.61$\pm$4.00 & 2.02 $\pm$ 2.30\\
            & \multirow{2}[0]{*}{\#Flops} & AE.Med. &0.63& 0.86 & 1.52 & 2.22 & 2.05 & 2.68 &  6.71  & 1.26\\
            &       & IQR   & 0.77 & 1.32 & 1.74 & 2.74 & 2.76 & 3.89 & 3.92 & 2.14 \\
        \midrule[0.5pt]
      \multirow{4}[0]{*}{EffiB0} & \multirow{2}[0]{*}{\#Para} & E.Med. & -0.17 & -0.05 & -0.06 & -0.06 & 0.67  & 2.16  & 4.86  & -0.03 \\
            &       &$\mu\pm\sigma$ & 0.88$\pm$1.92 & 0.73$\pm$0.76 & 1.11$\pm$1.09 & 1.55$\pm$1.31 & 2.72$\pm$3.54 & 3.54$\pm$4.27 & 8.10$\pm$8.89 & 1.30$\pm$2.24 \\
            & \multirow{2}[0]{*}{\#Flops} & AE.Med. & 0.54  & 0.57  & 0.85  & 1.31  & 1.89  & 2.27  & 4.86  & 0.78 \\
            &       & IQR   & 0.68  & 0.73  & 1.12  & 1.55  & 2.28  & 3.34  & 6.26  & 1.15 \\
        \bottomrule[1.5pt]
      \end{tabular}%
    \label{tab3}%
  \end{table*}%

\begin{figure}[!t]
    \centerline{\includegraphics[width=\columnwidth]{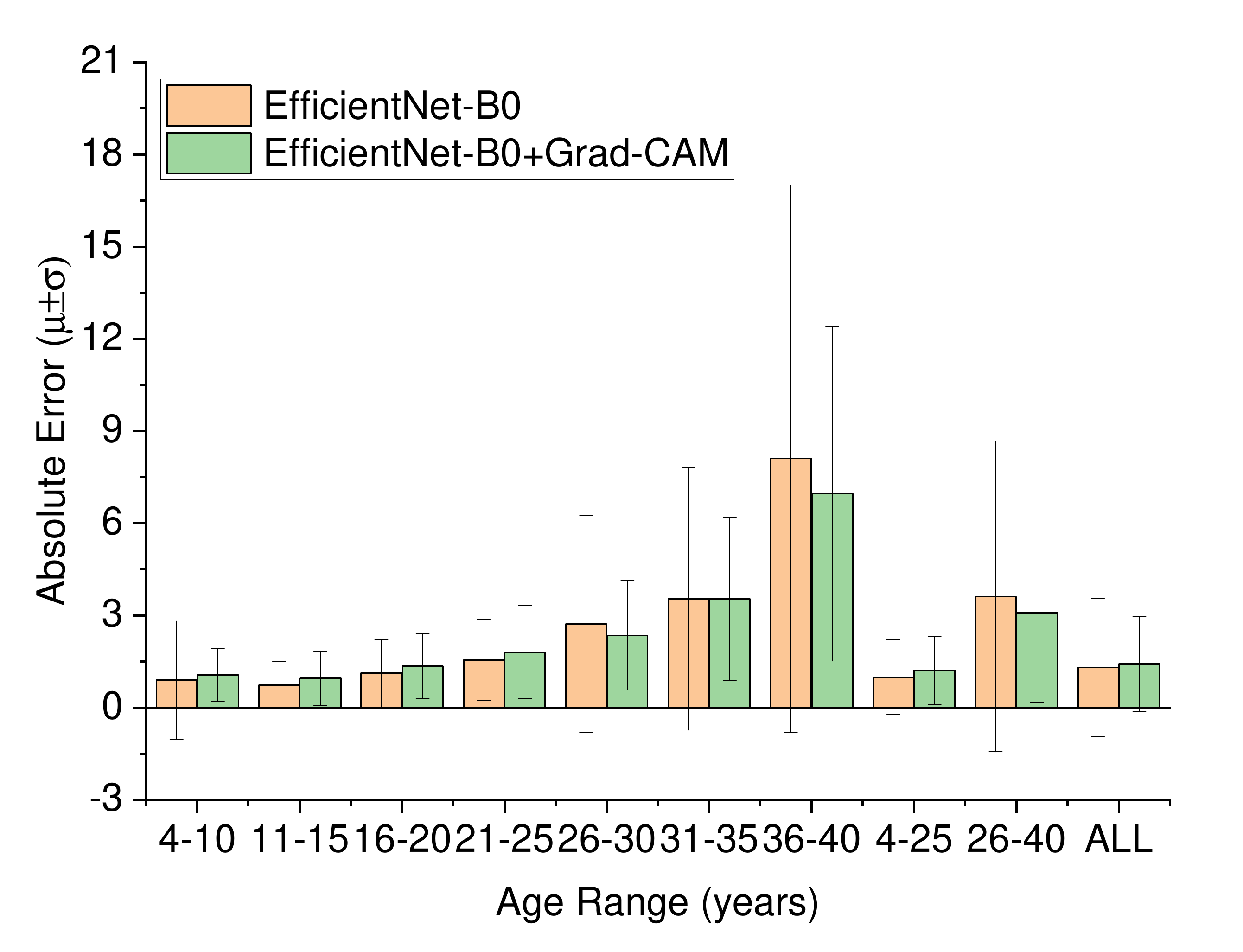}}
    \caption{Comparison of the performance of the basic network and the basic network constrained by the saliency graph}
    \label{fig compare-CAM}
\end{figure}

\begin{figure}[!t]
    \centerline{\includegraphics[width=\columnwidth]{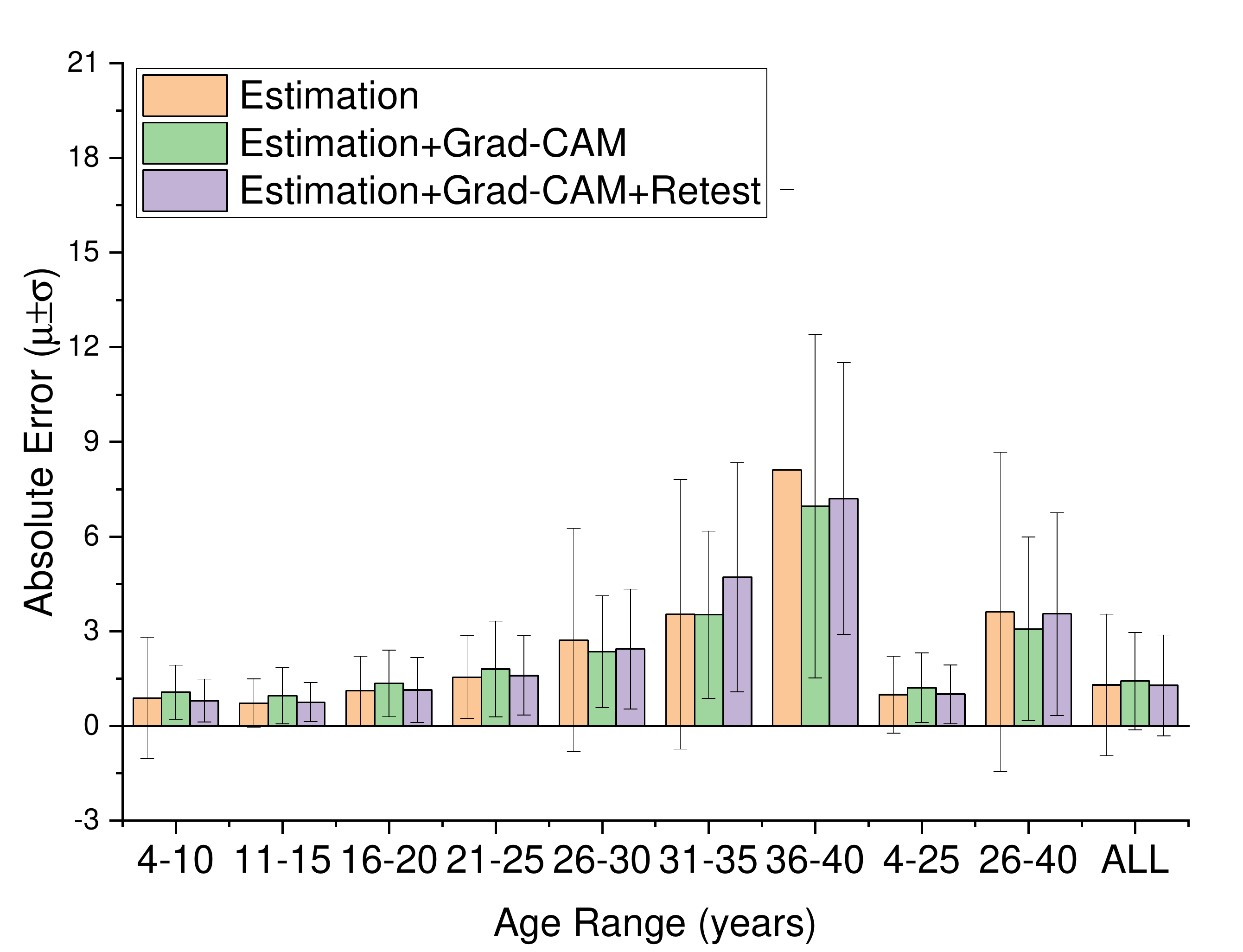}}
    \caption{Performance comparison of the basic network, 
    the network using saliency graph constraints and the network using saliency constraints and applying the retest mechanism}
    \label{fig compare-CAM-reset}
\end{figure}

\begin{figure*}[!t]
    \centerline{\includegraphics[width=\textwidth]{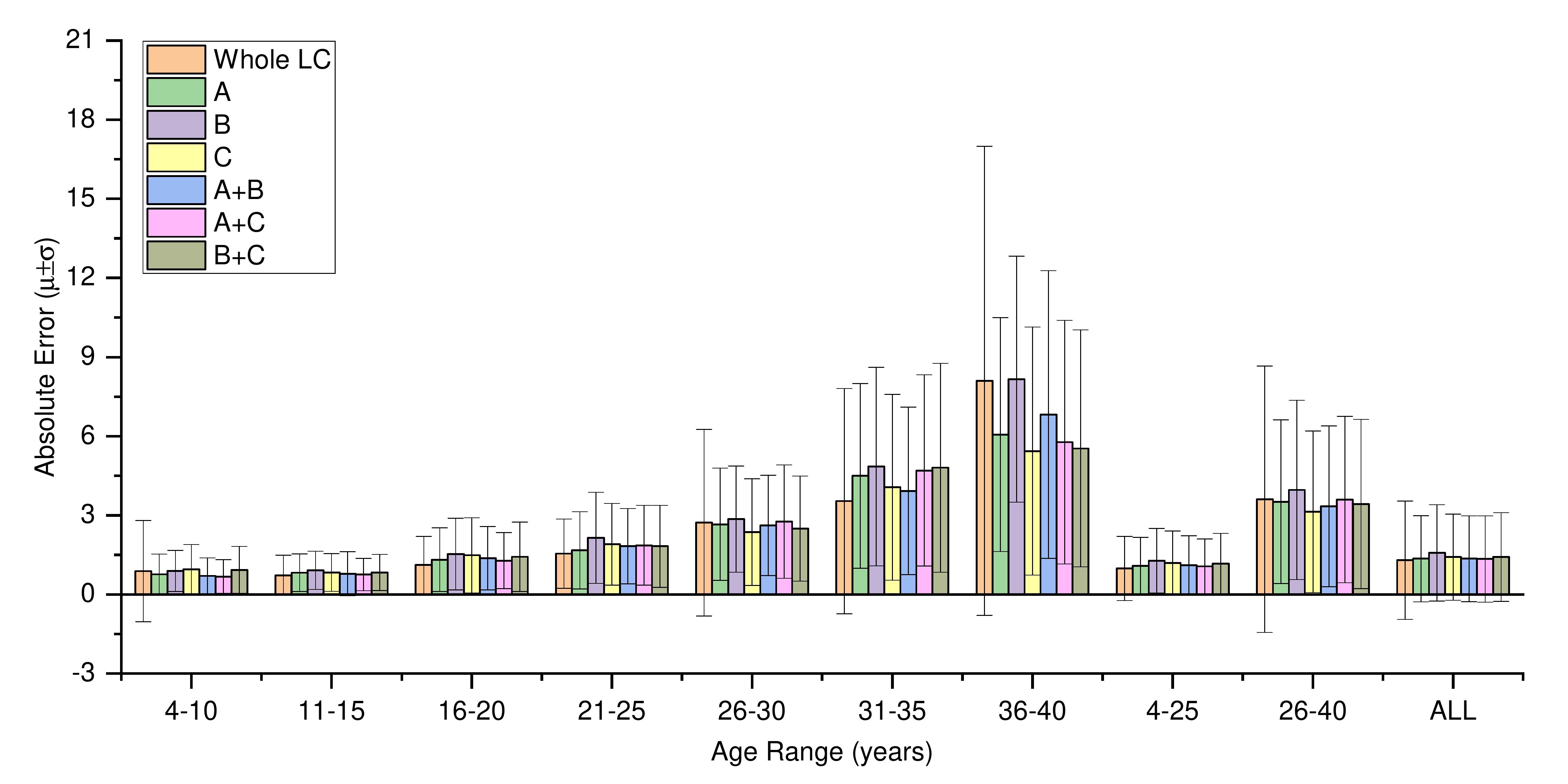}}
    \caption{When the input data is the entire LC image, 
    the three parts A, B and C are involved or missing, the performance comparison of age estimation}
    \label{fig compare-part}
\end{figure*}

\begin{figure}[!t]
    \centerline{\includegraphics[width=\columnwidth]{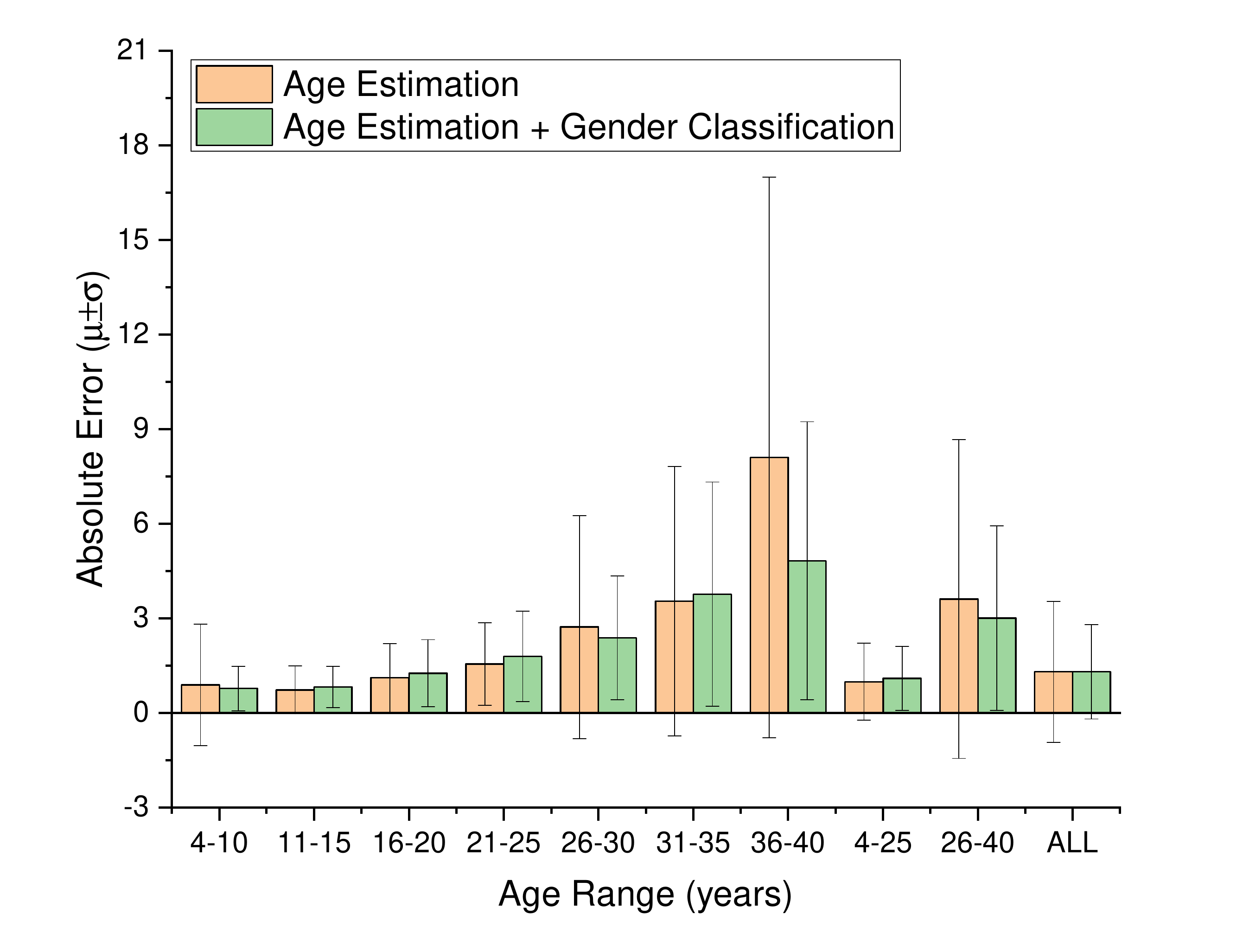}}
    \caption{Performance combination of age estimation and age estimation with gender classification}
    \label{fig compare-gender}
\end{figure}

\begin{figure}[!t]
    \centerline{\includegraphics[width=\columnwidth]{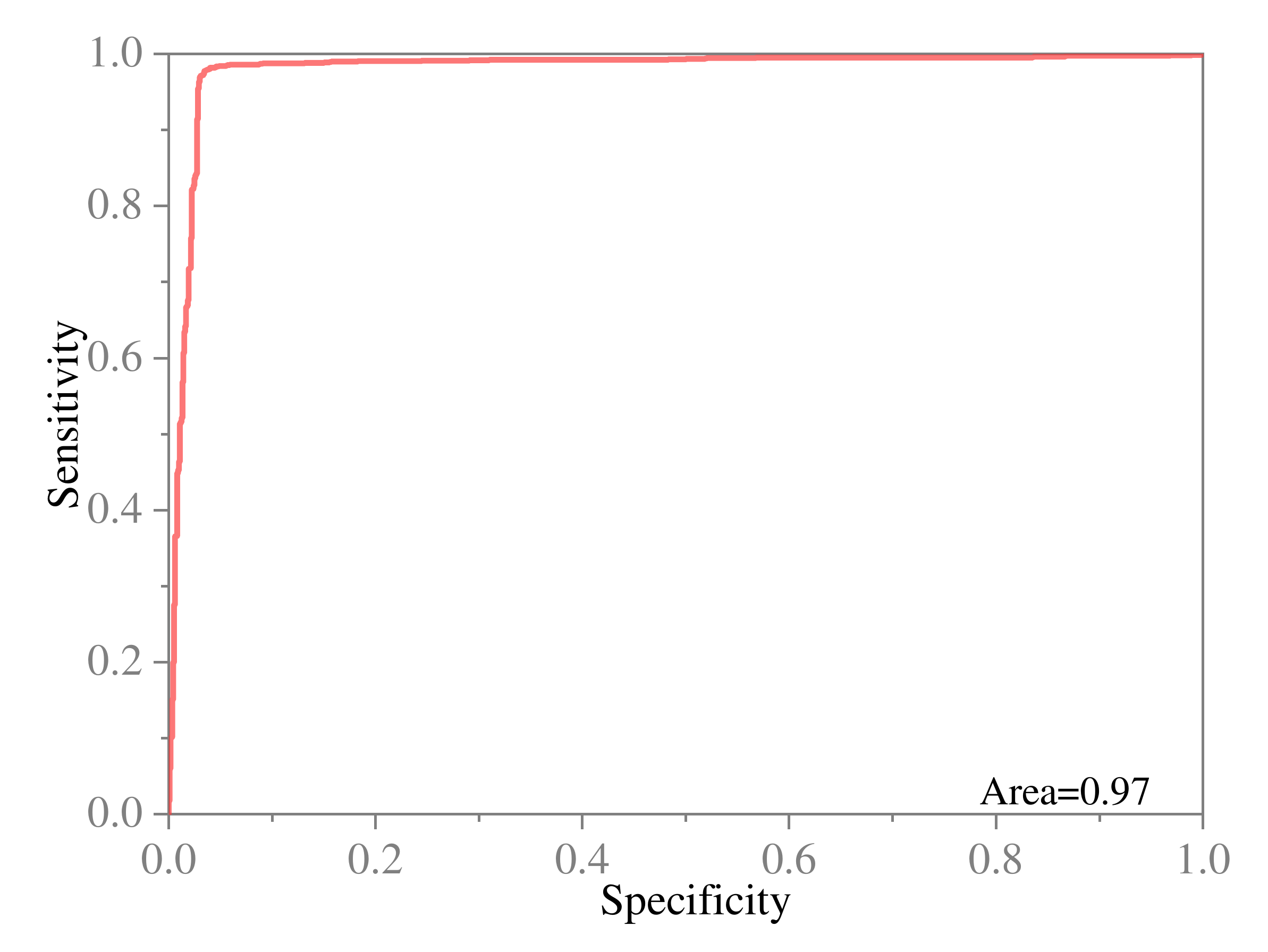}}
    \caption{ROC curve of basic network applied to gender classification. The AUC of ROC curve is 0.97.}
    \label{fig gender-roc}
\end{figure}

\section{RESULT}
\label{sec:RESULT}
As shown in Table. \ref{tab3},
for age estimation, 
Efficient-B0 performs far better than other networks in terms the average and dispersion of the prediction error except in the $36-40$ age range, 
and requires less memory and computation.
The overall mean absolute error(MAE) and standard deviation of the absolute error(SD) estimated by EfficientNet-B0 at all ages are $1.3$ (years) and $2.24$ (years), respectively.
The basic network performs best and worst in $11-15$ years old and $36-40$ years old, respectively,
which is highly correlated with the sample size of these two age ranges.
This is why we chosen EfficientNet-B0 as the basic network for our data set.
However, 
it can be seen from the results that the performance and stability of EfficientNet-B0 were not the best among the CNNs compared to it, 
for the samples aged from 26 to 40.

The performance of Grad-CAM being added to the input of the network is shown in Fig. \ref{fig compare-CAM}.
It can be seen that after adding Grad-CAM, 
compared to using only LC images as input, 
the MAE of our method on samples from 26 to 30 years old was significantly reduced.
More importantly, the addition of Grad-CAM reduced the overall SD, 
which means that the performance of the network is more stable, 
and the result is more reliable.

As shown in \ref{fig compare-CAM-reset} and \ref{tab:compare-all}, after the retest mechanism is added, 
the input of the network is different for samples of different ages, 
so the advantages of using Grad-CAM and not using Grad-CAM are combined.
Therefore, 
whether it is 4 to 25 years old or 26 to 40 samples, 
the MAE was not worse than the MAE that only uses LC images as input, 
and the overall MAE and SD was also reduced.
Compared with the better MAE, 
the optimization of SD is more worthy of attention, 
because when the mean value of absolute error is similar, 
the smaller standard deviation means that the performance of the network is more stable, 
which is of great significance in forensic practice.

The MAE and SD using only one part of the LC image or one part of the LC image missing for age estimation are compared in Fig \ref{fig compare-part}.
As can be seen in Fig. \ref{fig compare-part}, 
the performance of using only part A for age estimation is significantly better than using only one of the other two parts.
Naturally, 
if part A was missing, 
using the remaining two parts for age estimation had the worst performance.
However, 
the effrct of using only a certain part or missing a certain part on age estimation is not very significant, 
which shows that in forensic practice, 
only part of the anatomical structure can still be used for age estimation and is still credible.

As shown in Table \ref{tab:compare-all}, 
EfficientNet-B0 can be used for age estimation tasks and gender classification tasks simultaneously.
Moreover, 
due to the introduction of gender information, 
although only LC images are used as the input of the network, 
the performance of age estimation was better than other methods.
The MAE and SD of age estimation with gender classification are $1.24$ and $1.52$ respectively.
The performance comparison of age estimation with gender classification and the basic method is shown in Fig.\ref{fig compare-gender}.
The ROC for gender classification is shown in Fig. \ref{fig gender-roc}.
The accuracy and ROC-AUC of gender classification are $0.96$ and $0.97$ respectively.

\begin{table*}[htbp]
    \centering
    \caption{Comparison of performance in different age ranges and overall performance when different methods and input data are applied.
    EFFI.: Basic network, 
    EFFI.+CAM: Basic network with saliency graph constraints,
    EFFI.+CAM.+RTS.: Basic network with saliency graph constraint applied restest mechanism,
    PAR.A, PAR.B and PAR.C: Only part A, part B and part C are used as the input data of the basic network, respectively,
    PAR.A+B, PAR.A+C and PAR.B+C: Part C, Part B and Part A are missing in the input of the basic network, respectively,
    EFFI.+GEN.: The basic network performs age estimation and gender classification at the same time.}
    \begin{tabular}{m{75pt}<{\centering}
                      m{30pt}<{\centering}
                      m{30pt}<{\centering}
                      m{30pt}<{\centering}
                      m{30pt}<{\centering}
                      m{30pt}<{\centering}
                      m{30pt}<{\centering}
                      m{30pt}<{\centering}
                      m{30pt}<{\centering}
                      m{30pt}<{\centering}
                      m{32pt}<{\centering}
                      }
      \toprule[1.5pt]
      \diagbox[]{Method}{Age} & 4-10 & 11-15 & 16-20 & 21-25 & 26-30 & 31-35 & 36-40 & 4-25 & 26-40 & All \\
      \midrule[0.5pt]
      EFFI. & 0.88$\pm$1.92 & \textbf{0.73$\pm$0.76} & 1.11$\pm$1.09 & 1.55$\pm$1.31 & 2.72$\pm$3.54 & 3.54$\pm$4.27 & 8.10$\pm$8.89 & 1.00$\pm$1.22 & 3.61$\pm$5.06 & 1.30$\pm$2.24 \\[5pt]
      EFFI.+CAM. & 1.06$\pm$0.85 & 0.95$\pm$0.89 & 1.35$\pm$1.05 & 1.80$\pm$1.52 & 2.35$\pm$1.78 & 3.52$\pm$2.65 & 7.00$\pm$5.44 & 1.21$\pm$1.10 & 3.08$\pm$2.90 & 1.42$\pm$1.54 \\[5pt]
      EFFI.+CAM.+RTS. & 0.81$\pm$0.69 & 0.75$\pm$0.62 & 1.16$\pm$1.14 & 1.79$\pm$1.45 & 2.62$\pm$2.00 & 3.70$\pm$3.65 & \textbf{5.34$\pm$4.42} & 1.04$\pm$1.03 & 3.20$\pm$2.98 & 1.28$\pm$1.54 \\[5pt]
      PAR.A & 0.76$\pm$0.76 & 0.83$\pm$0.72 & 1.32$\pm$1.20 & 1.67$\pm$1.46 & 2.66$\pm$2.13 & 4.50$\pm$3.50 & 6.06$\pm$4.44 & 1.09$\pm$1.08 & 3.51$\pm$3.10 & 1.35$\pm$1.63 \\[5pt]
      PAR.B & 0.89$\pm$0.78 & 0.91$\pm$0.73 & 1.53$\pm$1.36 & 2.15$\pm$1.73 & 2.86$\pm$2.01 & 4.85$\pm$3.77 & 8.16$\pm$4.67 & 1.28$\pm$1.23 & 3.96$\pm$3.40 & 1.57$\pm$1.83 \\[5pt]
      PAR.C & 0.96$\pm$0.93 & 0.83$\pm$0.72 & 1.48$\pm$1.42 & 1.91$\pm$1.55 & 2.36$\pm$2.03 & 4.06$\pm$3.52 & 5.43$\pm$4.71 & 1.20$\pm$1.20 & 3.14$\pm$3.06 & 1.41$\pm$1.64 \\[5pt]
      PAR.A+B & 0.71$\pm$0.68 & 0.79$\pm$0.83 & 1.37$\pm$1.21 & 1.83$\pm$1.42 & 2.61$\pm$1.90 & 3.93$\pm$3.17 & 6.82$\pm$5.45 & 1.10$\pm$1.12 & 3.34$\pm$3.05 & 1.35$\pm$1.63 \\[5pt]
      PAR.A+C & \textbf{0.68$\pm$0.64} & 0.76$\pm$0.62 & 1.28$\pm$1.06 & 1.86$\pm$1.51 & 2.76$\pm$2.14 & 4.70$\pm$3.63 & 5.78$\pm$4.62 & 1.07$\pm$1.04 & 3.59$\pm$3.15 & 1.34$\pm$1.64 \\[5pt]
      PAR.B+C & 0.93$\pm$0.89 & 0.84$\pm$0.68 & 1.43$\pm$1.32 & 1.83$\pm$1.55 & 2.49$\pm$1.99 & 4.80$\pm$3.96 & 5.54$\pm$4.49 & 1.17$\pm$1.15 & 3.43$\pm$3.21 & 1.42$\pm$1.68 \\[5pt]
      EFFI.+GEN. & 0.70$\pm$0.65 & 0.77$\pm$0.79 & 1.19$\pm$0.96 & 1.73$\pm$1.40 & \textbf{2.22$\pm$1.73} & \textbf{3.50$\pm$3.09} & 6.43$\pm$5.78 & 1.03$\pm$1.02 & \textbf{2.93$\pm$3.00} & \textbf{1.24$\pm$1.52} \\
      \bottomrule[1.5pt]
    \end{tabular}%
    \label{tab:compare-all}%
\end{table*}%

\begin{figure*}[!t]
    \centerline{\includegraphics[width=\textwidth]{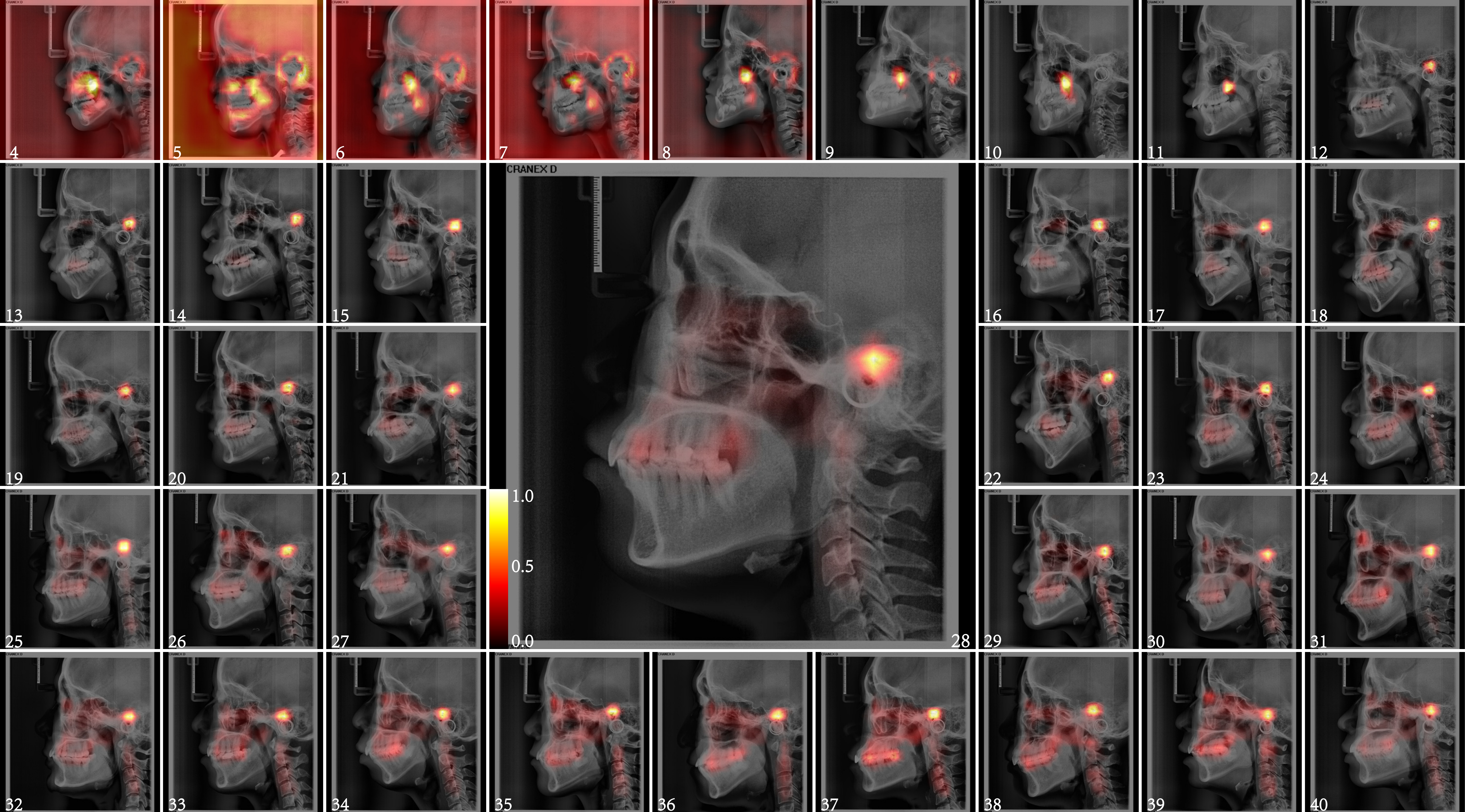}}
    \caption{Sample of saliency maps for each age.
    As shown by the gradient color bar, the brighter the region, 
    the more contribution to the age estimation}
    \label{fig cam-whole}
\end{figure*}

\begin{figure*}[!t]
    \centerline{\includegraphics[width=\textwidth]{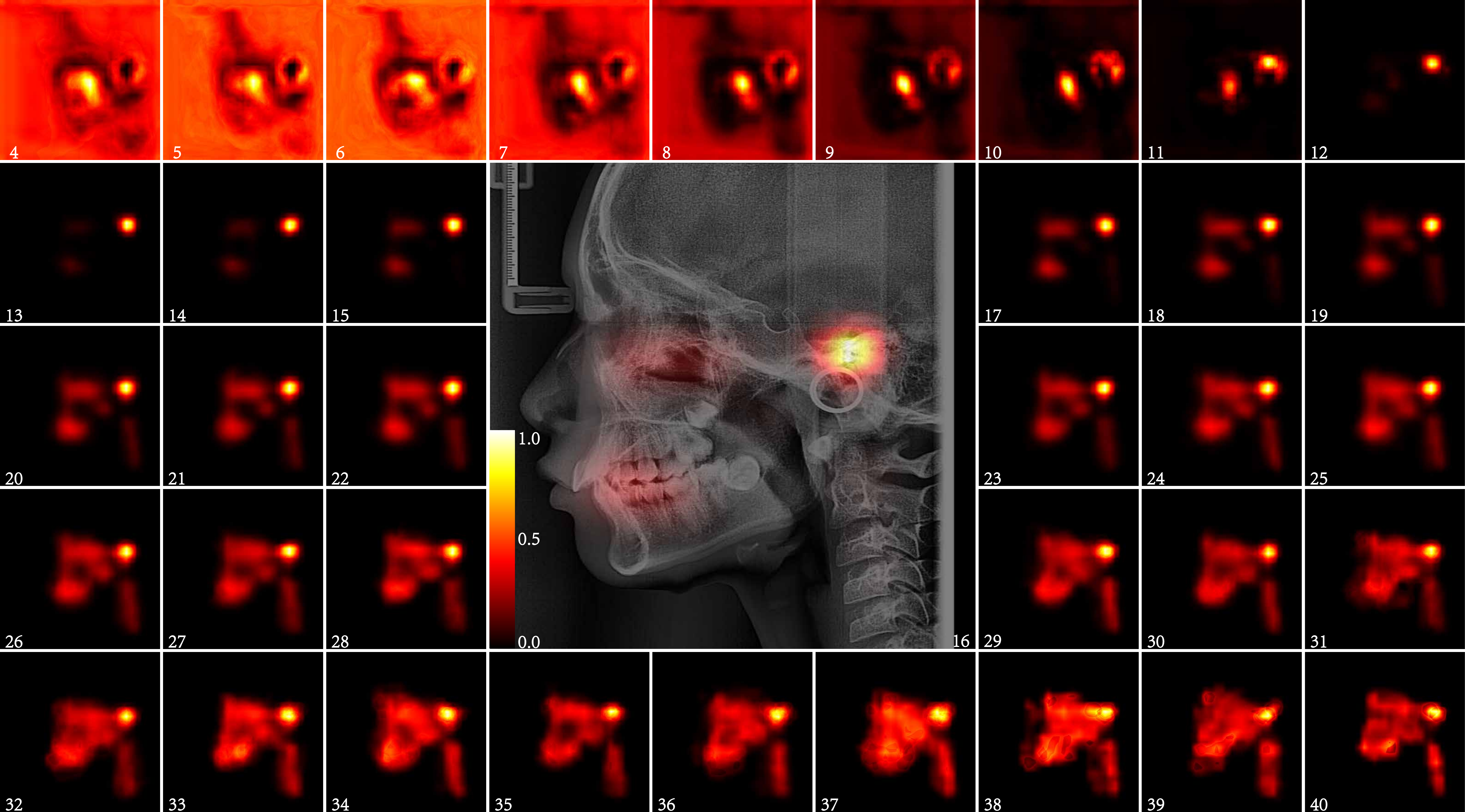}}
    \caption{The average of all saliency maps for each age. The average of all saliency maps for each age. 
    In order to show the relative position of the saliency map and the LC image, 
    a 28-year-old LC image and its corresponding average of saliency map are placed together.
    As shown by the gradient color bar, the brighter the region, 
    the more contribution to the age estimation}
    \label{fig cam-mean-whole}
\end{figure*}

\begin{figure*}[!t]
    \centerline{\includegraphics[width=\textwidth]{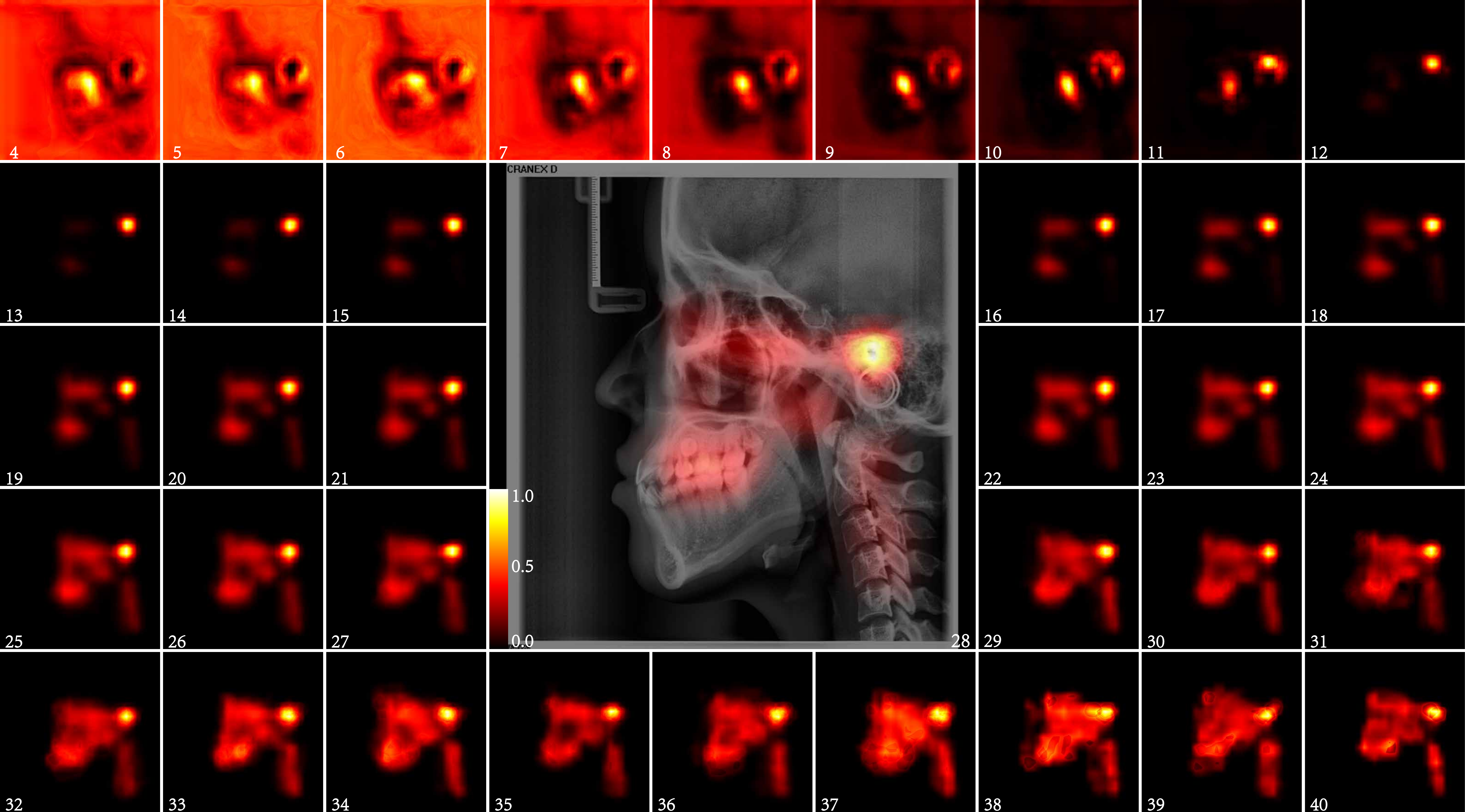}}
    \caption{The average of the saliency maps of all part A in each age. 
    In order to show the relative position of the saliency map and the LC image, 
    a Part A of 28-year-old image and its corresponding saliency map are placed together.}
    \label{fig cam-mean-tooth}
\end{figure*}

\begin{figure*}[!t]
    \centerline{\includegraphics[width=\textwidth]{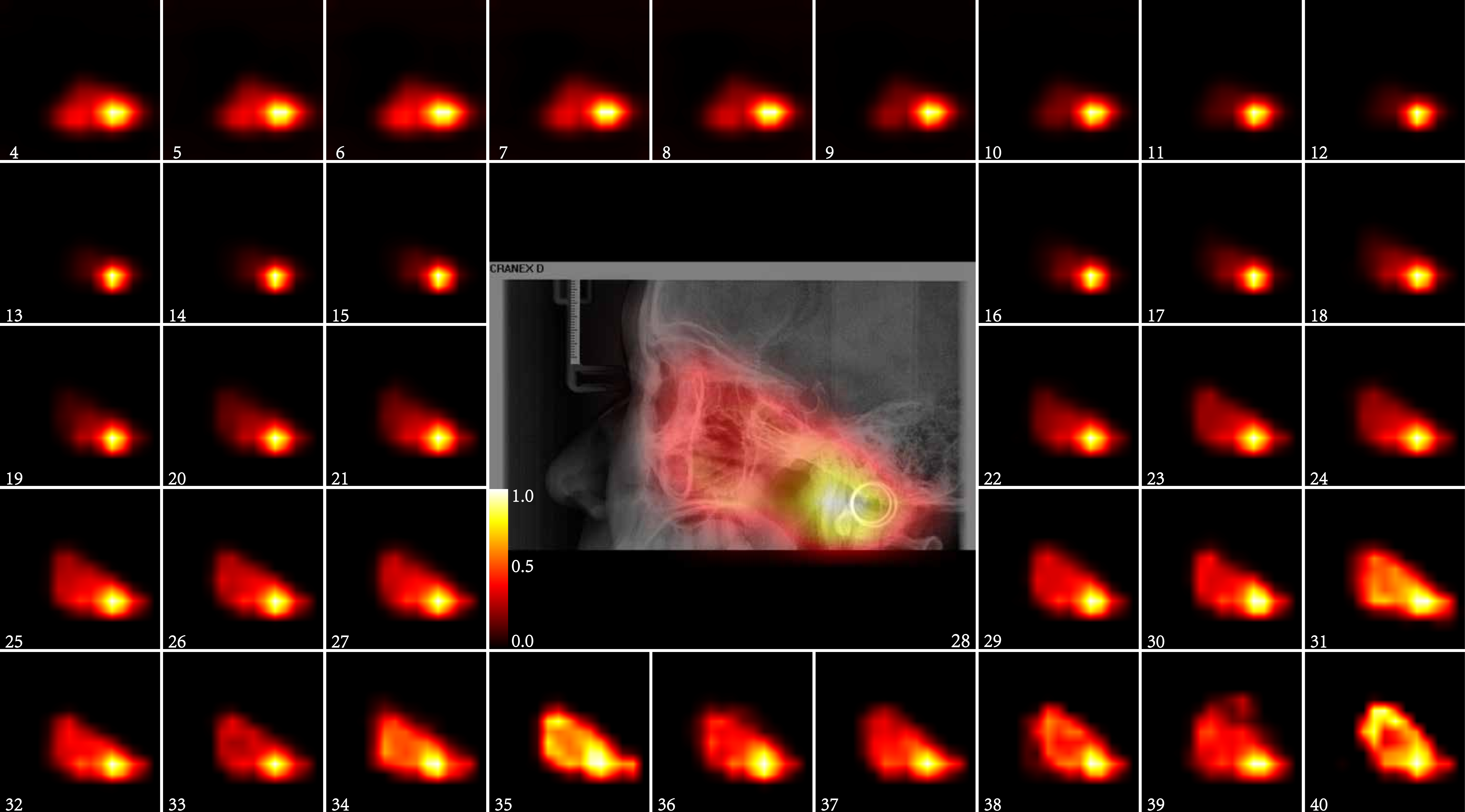}}
    \caption{The average of the saliency maps of all part B in each age. 
    In order to show the relative position of the saliency map and the LC image, 
    a Part B of 28-year-old image and its corresponding saliency map are placed together.}
    \label{fig cam-mean-top}
\end{figure*}

\begin{figure*}[!t]
    \centerline{\includegraphics[width=\textwidth]{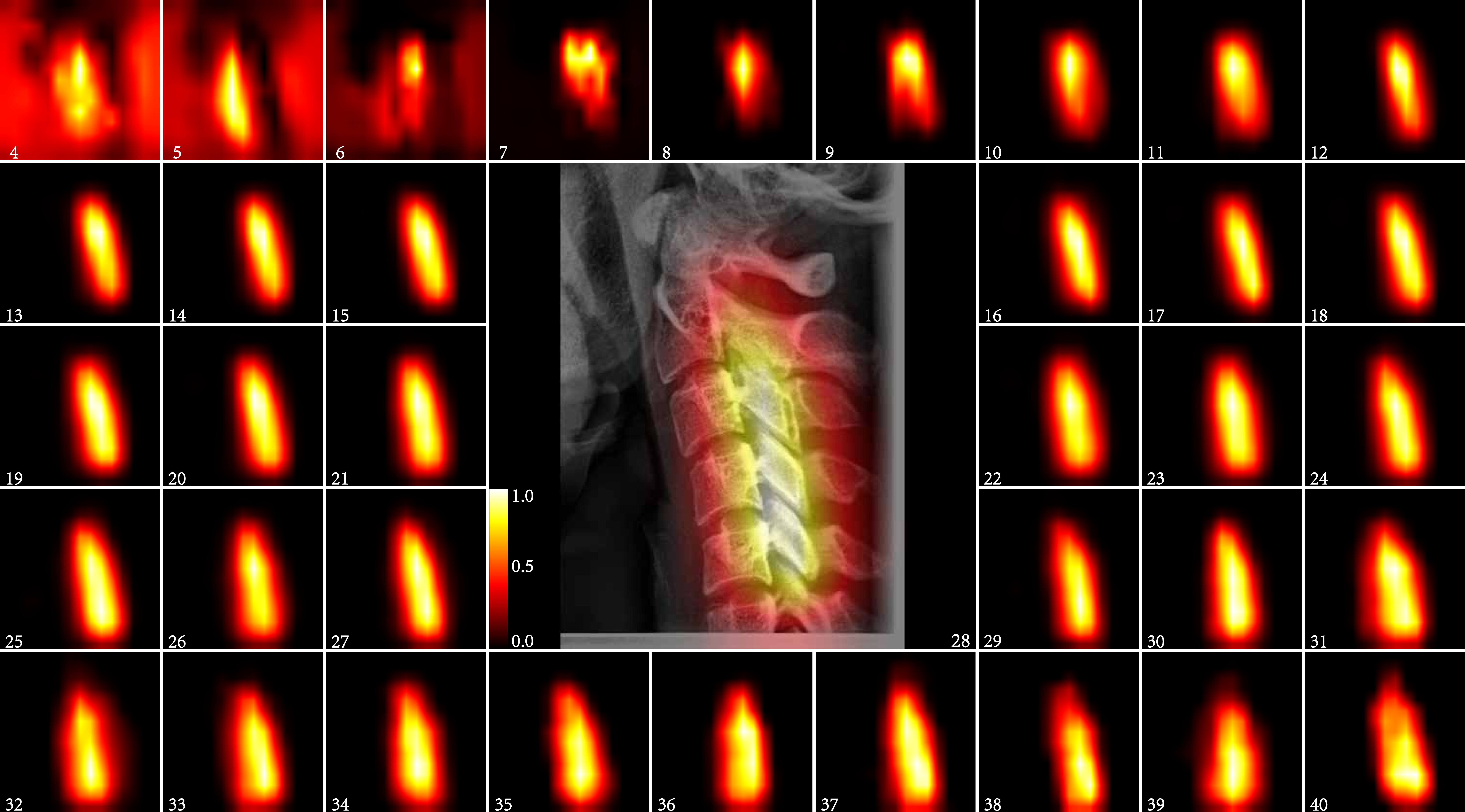}}
    \caption{The average of the saliency maps of all part C in each age. 
    In order to show the relative position of the saliency map and the LC image, 
    a Part C of 28-year-old image and its corresponding saliency map are placed together.}
    \label{fig cam-mean-vertebra}
\end{figure*}

\section{DISCUSSION}
\label{sec: DISCUSSION}
In recent years, 
due to the availability of medical datasets and the improvement of computer processing power, 
the application of deep learning algorithms in the field of computer vision has surged, 
and it has quickly become the preferred method for analyzing medical images. 
Convolutional neural networks have irreplaceable advantages in the field of image processing.
Its parameter sharing and local receptive fields allow the extraction of image features with fewer parameters without losing the spatial information of the image. 
Therefore, the convolutional neural network plays an irreplaceable role in the field of medical image processing and is one of the best deep learning models for researchers.

Deep learning has also brought about tremendous changes in age estimation research. 
Deep learning has incomparable advantages over traditional methods: 
1) It avoids the influence of subjective factors on age estimation; 
2) The accuracy and efficiency are higher than traditional manual methods; 
3) It is not labor-intensive.

Because the lateral cephalogram contains the cervical spine, 
all teeth and craniofacial bones,
as well as the craniofacial soft tissue. 
Therefore, 
it can simultaneously reflect the aging changes of these areas. 
So far, 
the study of age estimation using lateral radiographs is still a blank, 
so we choose lateral radiographs as the research object.

By comparing MAEs and SDs for each age group calculated by the four methods in this study(as shown in \ref{tab:compare-all}), 
we found that the MAE of the basic network is less than 1 year old when the age is less than 15 years old. 
This shows that the accuracy of the basic network is very high under the age of 15. With the increase of age, 
the MAE of basic network increases. 
Because the traditional methods using dental images can not accurately infer the age of over 25 years old, 
and the number of LC images of over 25 years old group is smaller than under 25 in our dataset, 
we calculated the MAEs of 4-25 age group and 26-40 age group respectively. 
The performance of the basic network is far superior to traditional methods, 
both in terms of the accuracy and efficiency of age estimation.

Grad-CAM can well describe the contribution of each area in the images to the output result and visualize it, 
so that we know which parts of the LC image are derived from the age estimation results and the contribution of each part. 
The interpretability of this method is crucial, 
which is also of great significance for us to understand the working mechanism of the network. 
When using EfficientNet-B0 with Grad-CAM to infer age, 
the MAE of the 26-40 age group is significantly lower than that of the basic network, 
with a decrease of $14.7\%$, 
which indicates that Grad-CAM can improve the accuracy of age inference for the age group with small sample size. 

The retest method combines the advantages of the former two methods. 
Compared with the basic network, 
the MAE of the 26-40 age group is significantly reduced($11.3\%$ reduction), 
and compared with the efficient net-b0 with Grad-CAM method, 
the MAE of the 4-25 age group is also significantly reduced($14\%$ reduction), 
The SDs of 4-25 and 26-40 age groups obtained by retset method are all smaller than that of the former two methods,
which reflects the stability of the age inference efficiency of retest method is higher than that of the former two methods.

When using EfficientNet-B0 for age estimation and gender classification at the same time, 
the MAE of the 26-40 age group is smaller than that of the retest method($8.4\%$ reduction), 
From the experimental results, 
the accuracy and stability of this method to infer age are the best among the methods used in this article,
but it requires that gender labels be available when training the network. The improvement of gender information for age estimation is mainly manifested in samples after the developmental period(26-40 age group). 
It also shows the influence of gender on age inference, especially for adults.

We take each year as an age group and select the LC images whose estimated age is closest to the actual age, 
and the saliency maps are generated to determine the key areas of different age groups that are most closely related to age inference, 
as shown in Fig. \ref{fig cam-whole} we have the following findings 
1) Before the age of nine, 
the skull and maxilla showed a strong correlation, 
which was consistent with the development stage of the skull and maxilla. 
2) At the age of 10-13, the maxilla,
mandible, cervical vertebra and skull showed a strong correlation, 
suggesting that the craniofacial and cervical vertebra had a rapid development and change in puberty. 
3) The saliency maps over 14 years old shows consistent salient area, which is divided into three areas, 
the tooth area, the cervical spine area and the craniofacial area without the teeth. 
4) With the increase of age, 
the change trend of the salient area is increasing, 
which indicates that the more tissues in LC image undergone aging changes with age. 
5) In the saliency maps over the age of 14 years, 
the most salient area related to age inference showed amazing consistency. 
It is located in the upper part of the external auditory canal. Due to the overlap of the left and right bone tissues of the LC image, 
we can not determine which part of the head the most salient area is. 
This requires subsequent 3D image research. 
The The average saliency maps of each year age group Fig. \ref{fig cam-mean-whole} can also well reflect the above findings.

By comparing MAEs and SDs of each parts for each age group (shown in Tab. \ref{tab:compare-all}), 
we found that among the three single parts in 4-25 age group, 
the accuracy of age estimation base on teeth part is the highest. In 26-40 age group, 
the cervical spine part is the highest. 
The results indicate that before and after the age of 25, the teeth and cervical spine have the strongest correlation with age inference. 
In the pairwise combination of three single parts,
in 4-25 age group the accuracy of age estimation base on part A+B(no the cervical spine) is the highest, 
followed by part A+C, in 26-40 age group, This once again illustrates the importance of teeth part.

The average saliency maps of each part were shown in Fig. \ref{fig cam-mean-tooth}, Fig.\ref{fig cam-mean-top} and Fig.\ref{fig cam-mean-vertebra}. 
The salient area of each part is consistent with that of the whole LC image. 
The location and shape of the salient area in the heat map of each year age group showed a high degree of consistency. 
The salient area of teeth part were mainly teeth and periodontal tissues, 
especially the upper posterior teeth, this should be related to the wear of the teeth and the ageing changes of the periodontal tissue.(Fig. \ref{fig cam-mean-tooth}) 
The salient area of the craniofacial part without the teeth were mainly midface(Fig.\ref{fig cam-mean-top}). 
Many scholars have conducted in-depth research on the aging changes of the orbit, 
and the volume of the orbit increases with age. 
The development of the maxilla is also a research hotspot, 
but the research on the aging of the maxilla in adults has not been involved. 
The saliency maps reminds us that it is necessary to conduct aging research on other organizational structures in the middle of the face. 
The salient area of the cervical spine part were all of cervical spines and intervertebral disc in the LC image. 
The morphological changes of the cervical spine are used to determining the pubertal growth spurt of adolescence[27-29]. 
The cervical spine was also used to infer age and gender [30-32] , 
but it is mainly used for children and adolescents, 
there is no relevant research using the cervical spine to infer the age of adults. 
The cervical spine consists of 7 vertebral bodies and intervertebral discs. 
After the development is complete, 
the structural changes of the cervical spine begin in middle age, 
but sometimes earlier[33]. I
ntervertebral disc degeneration begins at adolescence, 
and as it progresses, it can also leads to morphological alterations of the vertebral bodies. 
Cervical lordosis increased with age[34]. These changes are difficult to use by traditional methods of inferring age. 

In this study, we found that the accuracy of age estimation of cervical part is the highest among all parts in the 26-40 age group, 
which indicates that cervical region should be highly concerned, which is worthy of further study.

Since only orthodontic patients take LC images in oral clinical work, there are very few orthodontic treatment patients over 40 years old in China, 
and orthodontic patients under 4 years old rarely need lateral radiographs, 
so the sample age range of this study is 4-40 years old. That is to say, 
the methods used in this study cannot be used to infer age over 40. 
Because of the way the LC images were taken, the left and right craniofacial bones and teeth overlapped together, 
so the saliency regions in Grad-CAM cannot accurately describe the anatomic structures for age estimation, 
which need to be further compared and determined by the study of three-dimensional images in the future. 
Compared with the traditional artificial age estimation method, deep learning method is data-driven and needs a large number of data sets, 
which is a common defect of deep learning method. However, 
meta learning and small sample learning methods make it possible to use a small amount of data to achieve acceptable performance. 
This is also our future research interest.

\section{CONCLUSION}
\label{sec: CONCLUSION}
1.	In this paper, we proposed a novel method in which LC images are used for age estimation for the first time and has achieved relatively ideal results. 
Aiming at the problem of fewer samples and larger errors after the age of 25, 
the saliency map generated by Grad-CAM and the trained network are used to restrict the attention of the network, 
thereby improving the overall performance of the network, 
especially the performance of the samples after the age of 25.

2.	The saliency map was applied to visualize the contribution of each part of the image to the age estimation. 
The results found some new areas worthy of attention.

3.	Based on the saliency map, we also explored the performance of age estimation when different regions are involved or missing. 
The results proved that the information in the LC image is redundant. 
A single part can be used for age estimation, 
and its performance is comparable to that of the entire image.

4.	The gain of gender to age estimation is also verified in our work. 
When outputting gender and age at the same time, 
the accuracy and stability of the method inferring age is the best among the methods used in this article, 
especially the MAE of the age group over 25 years old is greatly reduced by 18.8\%.

\section*{Acknowledgment}
*

\section*{References and Footnotes}
*

\subsection{References}
\bibliographystyle{IEEEtran}
\bibliography{IEEEabrv, mylib}
\end{document}